\documentclass[twoside]{report}
\usepackage{amssymb, lj-igpl, amsmath, latexsym, enumitem, makeidx, epsfig, url, psfrag}
\Title[Expressing Properties in Second and Third Order Logic]{Expressing Properties in Second and Third Order Logic: Hypercube Graphs and SATQBF \footnote{Pre-print of article submitted to an special issue of the Logic Journal of the IGPL with selected papers from the 16th Brazilian Logic Conference.}}
\ShortAuthor{F. Ferrarotti, W. Ren and J. M. Turull Torres}
\LongAuthor{
\author{FLAVIO FERRAROTTI}
\address{School of Information Management, Victoria University of Wellington,
PO Box 600, Wellington 6140, New Zealand. E-mail:~flavio.ferrarotti@vuw.ac.nz}
\author{WEI REN}
\address{School of Engineering and Advanced Technology, Massey University,
Private Box 756, Wellington 6140, New Zealand. E-mail:~w.ren@massey.ac.nz}
\author{JOSE MARIA TURULL TORRES}
\address{ICTIC, Universidad de la Cuenca del Plata, Corrientes, Argentina and
Department of Informatics, Universidad Nacional de San Luis,
Ejercito de Los Andes 950, D5700HHW, San Luis, Argentina.
E-mail:~J.M.Turull@massey.ac.nz}} \Received{21 December 2012.}

\newtheorem{remark}{Remark}

\begin{document}
\begin{paper}

\begin{abstract}
It follows from the famous Fagin's theorem that all problems in NP are expressible in   
existential second-order logic ($\exists$SO), and vice versa. Indeed, there are well-known
$\exists$SO characterizations of NP-complete problems such as $3$-colorability, Hamiltonicity and clique. 
Furthermore, the $\exists$SO sentences that characterize those problems are simple and elegant.
However, there are also NP problems that do not seem to possess equally simple and elegant $\exists$SO characterizations. 
In this work, we are mainly interested in this latter class of problems.  
In particular, we characterize in second-order logic the class of hypercube graphs and the classes SATQBF$_k$ of satisfiable quantified Boolean formulae with $k$ alternations of quantifiers. We also provide detailed descriptions of the strategies followed to obtain the corresponding nontrivial second-order sentences. Finally, we sketch a third-order logic sentence that defines the
class SATQBF = $\bigcup_{k \geq 1}$SATQBF$_k$. The sub-formulae
used in the construction of these complex second- and third-order logic sentences, are good candidates to form part of a library of formulae. Same as libraries
of frequently used functions simplify the writing of complex
computer programs, a library of formulae could
potentially simplify the writing of complex second- and third-order queries,
minimizing the probability of error.
\end{abstract}

\section{Introduction}
\label{examples_HO}

Examples of second-order formulae expressing different properties of
graphs are fairly common in the literature. Classical
examples are $3$\emph{-colorability},
\emph{Hamiltonicity}, and \emph{clique} (see \cite{[Imm99],[Lib04]}
among others). These properties can be expressed by simple and
elegant second-order formulae. Likewise, there are graph properties
that can be expressed by simple and elegant third-order
formulae. One of those properties is that of being a \emph{hypercube graph}
(see \cite{[Fer08]}).
An \emph{$n$-hypercube graph} ${\bf Q}_n$, also called
an $n$-cube,  is 
an undirected graph whose vertices are binary $n$-tuples.
Two vertices of ${\bf Q}_n$ are adjacent iff they differ in exactly one bit.

The expressive power of third-order
logic is not actually required to characterize hypercube graphs, since 
they can be recognized in nondeterministic polynomial
time. Recall that by Fagin's theorem \cite{[Fag74]}, $\exists$SO captures NP.
Thus there are formulae in existential second-order logic
($\exists$SO) which can express this property. Nevertheless, 
to define the class of hypercube graphs in  
second-order logic is certainly more challenging than to define it
in third-order logic.

From an applied perspective, this indicates that it makes sense to investigate higher-order quantifiers in the context of database query languages. Despite the fact that most of the queries commonly used in the industry are in P, the use of higher-order quantifiers can potentially simplify the way in which many of those queries are expressed. 

Let SATQBF$_k$ denote the class of satisfiable quantified Boolean formulae with $k$ alternating blocks of quantifiers. From Fagin-Stockmeyer characterization of the polynomial-time hierarchy \cite{[Sto76]} and the fact that SATQBF$_k$ is complete for the level $\Sigma^p_k$ of that hierarchy \cite{[Wra76]}, it follows that for every $k \geq 1$, SATQBF$_k$ can be defined by a formula in the prenex fragment $\Sigma^1_k$ of second-order logic with $k$ alternating blocks of quantifiers. SATQBF$_k$ provides a prime example of a property (or query) whose expression in the language of second-order logic is possible but challenging. Indeed, it is not a trivial task to write a second-order logic sentence that evaluates to true precisely on those word models that represent sentences in SATQBF$_k$. As usual in finite model theory \cite{[EF99]}, the term word model refers here to a finite relational structure formed by a binary relation and a finite number of unary relations. By contrast, if we restrict our attention to quantified Boolean formulae in which the quantified free part is in conjunctive normal form and has exactly three Boolean variables in each conjunct, then the problem is expressible in monadic second-order logic provided that the formulae are encoded using a different kind of finite relational structures which include ternary relations~(see \cite{[Lib04]}).

Thus, on the one hand there are well-known NP-complete problems such as
$3$-colorability, Hamiltonicity and clique, that have corresponding well-known characterizations 
in $\exists$SO which are simple and elegant. Those characterizations
have in common that the existential second-order quantifiers
can be identified with the guessing stage of the NP algorithm, and that the
remaining first-order formula corresponds to the polynomial time
deterministic verification stage. On the other hand, there are well-known problems such as
hypercube graph (which can also be characterized in $\exists$SO) and SATQBF$_{k}$ (which can
be characterized in $\Sigma^1_k$) that do not appear to have a straightforward characterization in second-order logic, 
even if we consider the full second-order language.  

This observation prompted us to write second-order characterizations of hypercube graph and SATQBF$_k$. 
The resulting second-order sentence for hypercube graph can be found in \cite{[Rei08]}. The corresponding sentence for
 SATQBF$_k$ was included in \cite{[Rei11]}. Both sentences are complex and several pages long. 
In this article we present a detailed description of the strategies followed to write these
sentences. The sub-formulae used for the 
implementation of these strategies could be part of a future library of
second-order formulae. 
Same as libraries
of frequently used functions simplify the writing of complex
computer programs, a library of formulae could
potentially simplify the writing of complex second-order queries,
minimizing the probability of error.

The minimization of the probability of error constitutes an important objective in the context of this work, 
since given a query $q$ and a second-order formula $\varphi$, it is \emph{not} possible to formally prove whether $\varphi$ expresses $q$. 
For this reason, we make use of full second-order logic to present the characterizations of hypercube graph and SATQBF$_k$, even though its $\exists$SO and $\Sigma^1_k$ fragments, respectively,  already have the expressive power required for these tasks.
This has permitted us to write relatively clear and intuitive formulae as well as to follow a top-down strategy, similar to that commonly used in the development of computer programs, to further reduce the chance of error. 

If we consider the whole class SATQBF = $\bigcup_{k \geq 1}$SATQBF$_k$ of satisfiable quantified Boolean formulae, then the problem becomes PSPACE-complete. Since PSPACE can be captured by second-order logic extended with a transitive closure operator, and furthermore this logic is widely conjectured to be strictly more expressive than the standard second-order logic, the existence of a second-order logic characterization of this problem is unlikely. Thus, we decided to look for a characterization in third-order logic. Note that it is a well-known fact that third-order logic is powerful enough as to characterize every problem in PSPACE. 
We conclude the paper presenting a sketch of a third-order logic sentence that defines the
class SATQBF. That is, we present a strategy to write a third-order sentence  
 that evaluates to true precisely on those word models that represent sentences in SATQBF. 

We strongly believe that in many respects the descriptive approach
to Complexity is more convenient than the classical one. That is,
using formulae of some logic to study upper bounds in the time or
space complexity of a given problem, instead of Turing machines.
There are many different measures which can be taken on the formulae
that express a given problem such as quantifier rank, quantifier
blocks alternation, number of variables, number of binary
connectives, and arity of quantified relation variables. It has been proved 
that bounds on those measures impact on the 
expressive power of logics over finite models (see
\cite{[Lib04]}, \cite{[EF99]}, \cite{[Imm99]}). Furthermore, it is
rather obvious that all those measures are decidable, in contrast to
the use of Turing machines, where the usual measures relevant to
computation power such as time, space, treesize, and number of
alternations, are clearly undecidable. Regarding lower bounds there
are also several well studied and powerful techniques in Descriptive
Complexity which proved to be extremely useful in the last decades,
such as Ehrenfeucht-Fraisse games and their variations (see
\cite{[KT07]} in particular) and 0-1 Laws (again see \cite{[Lib04]}, 
\cite{[EF99]}, \cite{[Imm99]}).

Hence, it is important to learn how to build
formulae which are large, but still intuitive and clearly
understandable in a top down approach, in the same way that this is
important in the construction of algorithms in the classical
approach to Complexity, which are also clear and intuitive no matter
their size. The work reported in this article is to the authors'
knowledge one of the first steps in that direction.

In the next section, we introduce the necessary notation and
formally describe by means of a third-order logic sentence, the
class of hypercube graphs. In Section 3, we define in second-order
logic the basic arithmetic operations that we need for this work. We
describe the strategy used to characterize the class of hypercube
graphs in the language of second-order logic in Section 4. In
Section 5 we formally describe the problems SATQBF$_k$ and SATQBF,
and we consider their complexity. In Section 6, we explain in full
detail how to build for each $k \geq 1$, a second-order sentence that
expresses $SATQBF_k$. In Section 7 we explain how to build a third-order 
logic sentence which expresses SATQBF, and we give a sketch of
such formula. Finally in Section~8, we present some final considerations.

\section{Background}

We assume that the reader is acquainted with the basic concepts 
and the framework of Finite Model Theory \cite{[EF99],[Lib04]}. We use the notation from \cite{[Lib04]}.

We work on the vocabulary $\sigma = \{E\}$ of
graphs. An \emph{undirected graph} ${\bf G}$ is a finite relational structure of vocabulary
$\sigma$ satisfying $\varphi_1 \equiv \forall x y (E(x,y)
\rightarrow E(y, x))$ and  $\varphi_2 \equiv \forall x (\neg E(x, x))$. If we do not require
${\bf G}$ to satisfy neither $\varphi_1$ nor $\varphi_2$, then we speak of a \emph{directed graph} (or
\emph{digraph}). We denote as $V$ the domain of the structure ${\bf G}$, i.e., the set of vertices
of the graph ${\bf G}$. The edge relation of ${\bf G}$ is denoted as $E^{\bf G}$. 

By \emph{second-order logic} we refer to the logic
that is obtained when first-order
logic is extended with second-order variables which range over subsets and
relations defined over the domain, and quantification over such variables. As usual, we use uppercase
letters $X, Y, Z, \ldots$ to denote second-order variables and lower case
letters $x, y, z, \ldots$ to denote first-order variables. The arity of the
second-order variables that we use in our formulae is always clear from the context. 
See \cite{[Lib04]} or \cite{[EF99]} for a formal definition of second-order logic in the
context of finite model theory. We include an example of a second-order
formula that defines a simple graph property instead.  

\begin{example}\label{ex1}
  An undirected graph ${\bf G}$ is \emph{regular} if
  all its vertices have the same degree. It is well known that the class of
  regular graphs is not definable in first-order logic \cite{[EF99],[Imm99]}. In
  second-order logic, this class can be defined as follows:\\[0.2cm] 
$\exists A \big( \forall x \big( \exists B ( \mathrm{A1} \wedge \mathrm{A2}) \big) \big)$ where
\begin{itemize}[leftmargin=*]
\item A1 expresses ``$B$ is the set of vertices which are adjacent to $x$''.\\[0.1cm]
$\mathrm{A1} \equiv \forall z \big( B(z) \leftrightarrow E(x,z) \big)$ 
\item A2 expresses ``the sets $A$ and $B$ have the same cardinality''
with a formula stating that there is a bijection $F$ from $A$ to $B$.\\[0.1cm]
$\mathrm{A2} \equiv \exists F \forall x y z \big(\mathrm{A2.1} \wedge \mathrm{A2.2} \wedge \mathrm{A2.3} \wedge \mathrm{A2.4} \wedge \mathrm{A2.5} \big)$ where
\begin{itemize}
\item A2.1 means ``$F$ is a subset of $A \times B$''.\\
$\mathrm{A2.1} \equiv \big(F(x, y) \rightarrow A(x) \wedge B(y)\big)$
\item A2.2 means ``$F$ is a function''.\\
$\mathrm{A2.2} \equiv \big(F(x, y) \wedge F(x, z) \rightarrow y = z \big)$
\item A2.3 means ``$F$ is total''.\\
$\mathrm{A2.3} \equiv \big(A(x) \rightarrow \exists y (F(x, y))\big)$
\item A2.4 means ``$F$ is injective''.\\
$\mathrm{A2.4} \equiv \big(F(x, z) \wedge F(y, z) \rightarrow x = y\big)$ 
\item A2.5 means ``$F$ is surjective''.\\
$\mathrm{A2.5} \equiv \big(B(y) \rightarrow \exists x (F(x, y))\big)$
\end{itemize}
\end{itemize}
\end{example}

We say that  a  sentence $\varphi$ expresses a Boolean 
query $q$ (or property) over finite relational structures of vocabulary 
$\sigma$, if for every finite relational structure ${\bf G}$ of vocabulary
$\sigma$, $q({\bf G})=\mathrm{true}$ iff
${\bf G} \models \varphi$. For instance the sentence in Example~\ref{ex1} expresses 
the Boolean query: Is ${\bf G}$ a regular graph?
We denote by
$Mod(\varphi)$ the class of finite $\sigma$-structures ${\bf G}$ such
that ${\bf G} \models \varphi$.  A class of finite $\sigma$-structures
${\cal C}$ is \emph{definable} in a logic $\cal L$, if ${\cal C} =
Mod(\varphi)$ for some ${\cal L}$-sentence $\varphi$ of vocabulary $\sigma$. For instance
  the class of regular graphs is definable in second-order logic, as shown by the formula given in Example~\ref{ex1}. 

Next, we define the class of hypercube graphs using a relatively simple and
elegant formula in \emph{third-order logic}. This logic extends second-order
logic with third-order variables which range over subsets and relations 
defined over the powerset of the domain, and quantification over such variables. We use uppercase
calligraphic letters ${\cal X}, {\cal Y}, {\cal Z}, \ldots$ to denote
third-order variables. A formal definition of higher-order logics in the
context of finite model theory can be found in \cite{[HT06]} among others.  
\begin{example}\label{n_cube}
  An \emph{$n$-hypercube} (or \emph{$n$-cube} for short) ${\bf Q}_n$ can be
defined as an undirected graph whose vertices are all the binary $n$-tuples.
  Two vertices of ${\bf Q}_n$ are adjacent iff they differ in exactly one bit.
  A $1$-cube ${\bf Q}_1$, a $2$-cube ${\bf Q}_2$ and a $3$-cube ${\bf
  Q}_3$ are displayed in Figure~2.1.
  \begin{figure}\label{cube}
  \centering\epsfig{file=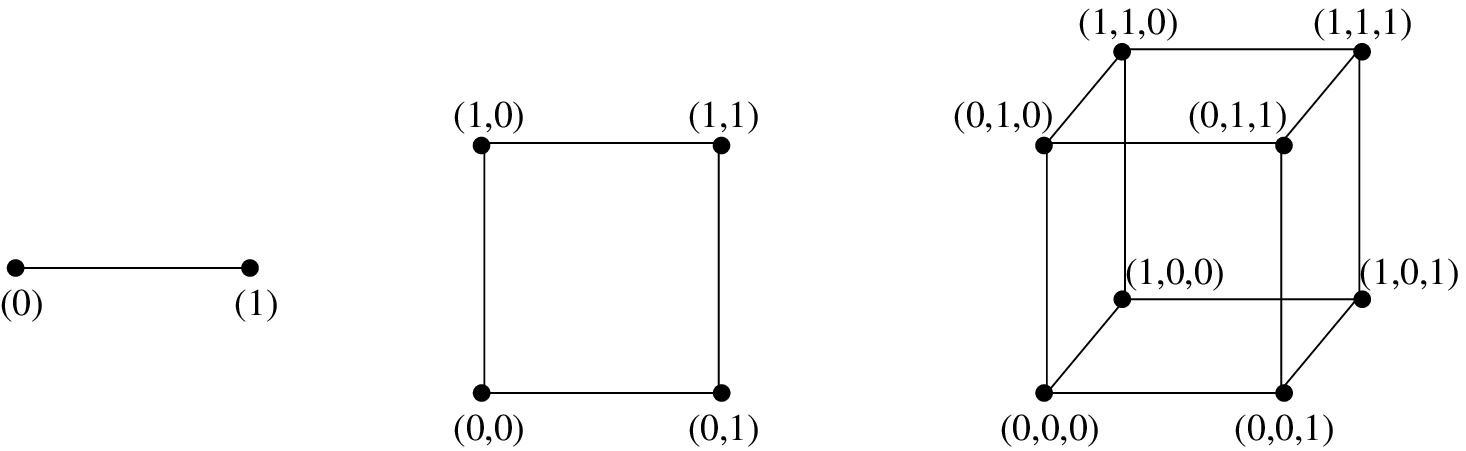,height=2.8cm}
  \begin{center}      
    Figure~2.1  
  \end{center}
  \end{figure}
  
  We can build an $(n+1)$-cube ${\bf Q}_{n+1}$ starting with two
  isomorphic copies of an $n$-cube ${\bf Q}_n$ and adding edges between
  corresponding vertices. Using this fact, we can define in
  third-order logic the so called class of \emph{hypercube graphs}, as
follows:\\[0.2cm]    
\noindent
$\exists {\cal C} \exists {\cal O} \big(\mathrm{A1} \wedge \mathrm{A2} \wedge \forall G_1 \forall G_2 \big(({\cal C}(G_1) \wedge
{\cal C}(G_2) \wedge \mathrm{A3}) \rightarrow \mathrm{A4} \big) \wedge \mathrm{A5} \wedge \mathrm{A6} \big)$ where
\begin{itemize}[leftmargin=*]
\item A1 expresses ``${\cal C}$ is a class of undirected graphs''.
\item A2 expresses ``${\cal O}$ is a total order on ${\cal C}$''.
\item A3 expresses ``$G_1$ is the immediate predecessor of $G_2$ in the order ${\cal O}$''.
\item A4 expresses ``$G_2$  can be built from two isomorphic copies of $G_1$ by adding edges between the corresponding vertices''.
\item A5 expresses ``the first graph in the order ${\cal O}$ is a $Q_1$''.
\item A6 expresses ``the last graph in the order ${\cal O}$ is the input graph''.
\end{itemize}
In turn, we can express A4 as follows:\\[0.2cm]
$\exists F_1 \exists F_2 \big(\mathrm{A4.1} \wedge \mathrm{A4.2} \wedge \mathrm{A4.3} \wedge \forall x (x \in dom(G_1) \rightarrow \mathrm{A4.4} ) \wedge$\\[0.1cm] 
\hspace*{1.2cm} $\neg \exists x y (x, y \in dom(G_1) \wedge x \neq y \wedge \mathrm{A4.5} )\big)$ where 
\begin{itemize}[leftmargin=*]
\item A4.1 expresses ``$F_1$ and $F_2$ are injective and total functions from $dom(G_1)$ to $dom(G_2)$''.
\item A4.2 expresses ``the ranges of $F_1$ and $F_2$ form a partition of $dom(G_2)$''.
\item A4.3 expresses ``$F_1$ and $F_2$ are isomorphisms from $G_1$ to the sub-graphs of $G_2$ induced by the ranges of $F_1$ and $F_2$, respectively''.
\item A4.4 expresses ``there is an edge in $G_2$ which connects $F_1(x)$ and $F_2(x)$''.
\item A4.5 expresses ``there is an edge in $G_2$ which connects $F_1(x)$ and $F_2(y)$''.
\end{itemize}
Note that, if there is an edge $(a,b)$ in $G_2$ such that $a$ belongs to the
range of $F_1$ and $b$ belongs to the range of $F_2$, or vice versa, then either $F^{-1}_1(a) =
F^{-1}_2(b)$ or $F^{-1}_1(b) = F^{-1}_2(a)$.  

The missing logic formulae in this example are left as an exercise for the reader.     
\end{example}


The property of a graph being an $n$-cube for some $n$, is known to be in NP.
A nondeterministic Turing machine can decide in polynomial time whether an
input structure ${\bf G}$ of the vocabulary $\sigma$ of graphs is an
hypercube, by simply computing the following steps:
\begin{enumerate}[label=\roman*.]
\item Compute the logarithm in base $2$ of the size $n$ of the domain of the input
structure ${\bf G}$ which must be a positive integer;
\item Guess a sequence $s_1, \ldots, s_n$ of $n$ binary strings, each of length $\log_2 n$;
\item Check in polynomial time that all binary strings are unique, that the sequence contains all binary strings of length $\log_2 n$ and that,
for some ordering $a_{s_1}, \ldots, a_{s_n}$ of the nodes in $V$, a string $s_i$ differs 
from a string $s_j$ in exactly $1$ bit iff there is an edge
$(a_{s_i}, a_{s_j}) \in E^{\bf G}$.   
\end{enumerate}

Thus, as we mentioned in the introduction, the full expressive power of third-order
logic is not actually needed to characterize the class of hypercube graphs. In fact, there is a
formula in $\exists$SO which can express this property. Recall that by Fagin's 
theorem \cite{[Fag74]}, $\exists$SO captures $NP$. However, it is very
unlikely that there is a formula in second-order logic, not to mention in 
$\exists$SO, that expresses the property
in a way which is as intuitive and simple as in the example above.

\section{Arithmetic in Second-Order Logic}

We define in this section the basic arithmetic operations of addition,
multiplication and exponentiation in second-order logic over finite structures. We encode initial
segments of natural number as finite relational structures by using linear
digraphs. Let $\bf G$ be a linear digraph. The first (root) element of the domain in the order determined by
the edge relation $E^{\bf G}$ represents the $0$, the second element in this
order represents the $1$, the third element represents the $2$ and so on. 
Since in a linear digraph, $E^{\bf G}$ is the successor relation, for clarity we use $\mathrm{succ}(x,y)$ to denote $E(x,y)$. We also use $x = n$ where $n > 0$ to denote the formula of the form
\[\exists y (\mathrm{succ}(y, x) \wedge \exists x (\mathrm{succ}(x, y) \wedge \exists y (\mathrm{succ}(y, x) \wedge \cdots \wedge \varphi) \cdots ))\]
with $n$ nested quantifiers and $\varphi \equiv \neg \exists x (\mathrm{succ}(x, y))$ if $n$ is odd or $\varphi \equiv \neg \exists y (\mathrm{succ}(y,x))$ if $n$ is even. Likewise, $x = 0$ denotes $\neg \exists y (\mathrm{succ}(y,x))$.
We assume a total order
$\leq$ of the nodes in $V$ such that $x \leq y$ iff there is a path from
$x$ to $y$ in $\bf G$ or $x = y$. This total order is easily definable in
second-order logic.   
\begin{figure}\label{sum}
\centering
\psfrag{0}{\scriptsize$0$}
\psfrag{1}{\scriptsize$1$}
\psfrag{2}{\scriptsize$2$}
\psfrag{succ}{\scriptsize$\mathrm{succ}(x)$}
\psfrag{label}{\scriptsize$y$ (if $x$ and $y$ are not $0$)}
\psfrag{F}{\scriptsize$F$}    
\psfrag{x}{\scriptsize$x$}
\psfrag{y}{\scriptsize$y$}
\psfrag{z}{\scriptsize$z=x+y$}
\epsfig{file=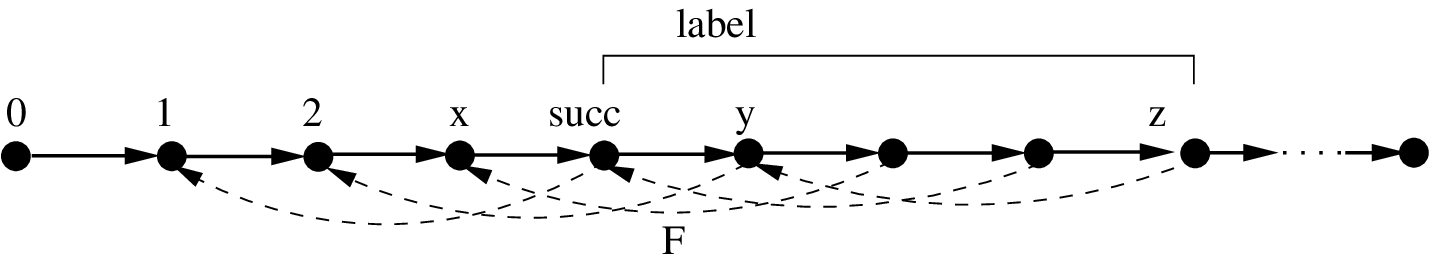,height=1.6cm}
\begin{center}   
 Figure~3.1: Addition  
\end{center}
\end{figure}

Let us start by defining the operation of \emph{addition}. The strategy is depicted in
Figure~3.1  in which we show the result $z$ of
adding $x$ and $y$ along a linear graph. 

The predicate $\mathrm{sum}(x, y, z)$, which is true iff $z
= x + y$, can be defined in second-order logic as follows.\\[0.2cm]   
\noindent
$\big(x = 0 \wedge z = y\big) \vee \big(y = 0 \wedge z = x\big) \vee$\\[0.1cm]
$\big(x \neq 0 \wedge y \neq 0 \wedge \exists F \big(\mathrm{A1} \wedge F(z, y) \wedge \exists x' y' (\hspace{0.1cm} \mathrm{succ}(x, x') \wedge
F(x', y') \wedge y' = 1) \wedge$ \\[0.1cm]
\hspace*{3cm} $\forall x' y' x'' y'' ((\hspace{0.1cm} \mathrm{succ}(x', y')
\wedge F(x', x'') \wedge F(y', y'')) \rightarrow \hspace{0.1cm}
\mathrm{succ}(x'', y'')) \big) \big)$\\[0.2cm]
where A1 expresses ``$F$ is an injective function with domain $\{n \in V \hspace{0.1cm}|\hspace{0.1cm} \mathrm{succ}(x) \leq n \leq z\}$''.
It is an easy and supplementary task to write the actual formula corresponding to A1. For the sake of clarity, we avoid this kind of supplementary details from now on.

The next
arithmetic operation that we define is \emph{multiplication}. The strategy is
depicted in Figure~3.2 in which we show the result $z$ of $x$ times
$y$. Each of the nodes in the subset $S = \{2,
\ldots, x\}$ can be considered as a root of a different ordered tree in a forest. Each
root node in the forest has $y$ children and the result $z$ is the last child
of node $x$.       
\begin{figure}\label{times}
\psfrag{0}{\scriptsize$0$}
\psfrag{1}{\scriptsize$1$}
\psfrag{2}{\scriptsize$2$}
\psfrag{y1}{\scriptsize$y(1)$}
\psfrag{y2}{\scriptsize$y(2)$}
\psfrag{yx1}{\scriptsize$y(x-1)$}
\psfrag{S}{\scriptsize$S$}    
\psfrag{x}{\scriptsize$x$}
\psfrag{y}{\scriptsize$y$}
\psfrag{z}{\scriptsize$z=x \times y$}
\centering\epsfig{file=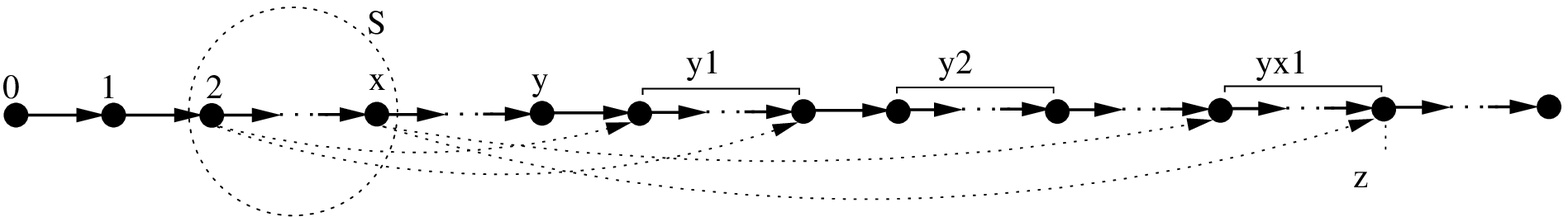,height=1.6cm}
\begin{center}      
 Figure~3.2: Multiplication  
\end{center}
\end{figure}

The predicate $\mathrm{times}(x,y,z)$, which is true if $z = x
\times y$, can be defined in second-order logic as follows. \\[0.2cm]
$(x = 1 \wedge y \neq 0 \wedge z = y) \vee (y = 1 \wedge x \neq 0 \wedge z =
x) \vee ((x = 0 \vee y = 0) \wedge z = 0) \vee$\\[0.1cm]
$\big(x \neq 0 \wedge y \neq 0 \wedge x \neq 1 \wedge y \neq 1
\wedge$\\[0.1cm]
\hspace*{0.1cm} $\exists S \big( \forall u \big( \big((2 \leq u \wedge u \leq x) \rightarrow [\exists
y'(S(u, y')) \wedge $\\[0.1cm]
\hspace*{4.4cm} $\forall x' y' ( (x' \leq y' \wedge x' \neq y' \wedge S(u, x')
\wedge S(u, y')) \rightarrow$\\[0.1cm]
\hspace*{5.4cm} $\neg \exists z' (x' \leq z' \wedge z' \leq y'
\wedge \neg S(u, z')))\wedge$\\[0.1cm]
\hspace*{4.4cm} $\exists F (\mathrm{A1}) \wedge \mathrm{A2} \wedge \mathrm{A3} \wedge \mathrm{A4}]\big) \wedge$\\[0.1cm]
\hspace*{1.4cm} $\mathrm{A5} \wedge \forall u v (S(u, v) \rightarrow (2 \leq u \wedge u \leq x))\big)\big)$ where
\begin{itemize}[leftmargin=*]
\item A1 expresses ``$F$ is a bijection from $\{n \in V \hspace{0.1cm}|\hspace{0.1cm} S(u, n) \}$ to $\{n \in V \hspace{0.1cm}|\hspace{0.1cm} 1 \leq n \leq y\}$, which means that the output degree of $u$ is $y$''.
\item A2 expresses ``if $u = 2$ then the first child of $u$ is $\mathrm{succ}(y)$''.
\item A3 expresses ``if $u = x$ then the last child of $u$ is $z$''.
\item A4 expresses ``if $u \neq 2$ then $\mathrm{succ}(c_{u-1}, c_u)$ for $c_{u-1}$ the last child of $u-1$ and $c_u$ the first child of $u$''.
\item A5 expresses ``the input degree of every node in $S$ is $\leq 1$''.
\end{itemize}

Finally, we need to define the arithmetic operation of \emph{exponentiation} in second-order logic. In this case, the strategy is depicted in
Figure~3.3. Note that, the first node in the linear digraph is 
$x^1$, the second node is $x^2$, and so on till node $y$-th (the final
node) which is $x^y$. 
\begin{figure}\label{exp}
\psfrag{x}{\scriptsize$x$}
\psfrag{x2}{\scriptsize$x \times x$}
\psfrag{x3}{\scriptsize$x \times x^2$}
\psfrag{z}{\scriptsize$z = x \times x^{y-1}$}
\centering\epsfig{file=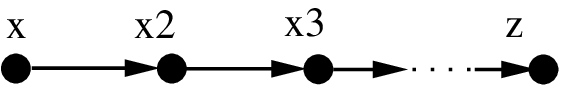,height=0.6cm}
\begin{center}      
 Figure~3.3: Exponentiation  
\end{center}
\end{figure}

The predicate $\mathrm{exp}(x,y,z)$, which is true if $z = x^y$, can be
defined in second-order logic as follows. \\[0.2cm]
$(x \neq 0 \wedge y = 0 \wedge z = 1) \vee (y = 1 \wedge z = x) \vee (x = 1 \wedge z = 1)
\vee$\\[0.1cm]
$\big(x \geq 2 \wedge y \geq 2 \wedge \exists V' E' \big(\mathrm{A1} \wedge \exists F (\mathrm{A2}) \wedge$ \\[0.1cm]
\hspace*{1cm} $\forall u (\neg V'(u) \vee (u = x \vee \exists x' (E'(x', u) \wedge \mathrm{times}(x, x',
u))))\big)\big)$ where    
\begin{itemize}[leftmargin=*]
\item A1 expresses ``$(V', E')$ is a linear digraph whose first (root) node is $x$ and whose last (leaf) node is $z$''.
\item A2 expresses ``$F$ is a bijection from $V'$ to $\{1, \ldots, y\}$, i.e., $|V'| = y$''.
\end{itemize}

\section{Hypercube Graph in Second-Order Logic}

We describe in this section two different strategies to define in second-order logic the
class of hypercube graphs. The first strategy is based in the usual definition of Hypercube graph which identifies the nodes of the graph with binary strings. This definition was explained and expressed by means of a third-order logic formula in Example~\ref{cube}. The second strategy is based in the following definition: An $n$-hypercube graph is a graph with $2^n$ nodes, which correspond to the subsets of a set with $n$ elements. Two nodes labelled by subsets $S_i$ and $S_j$ are joined by an edge if and only if $S_i$ can be obtained from $S_j$ by adding or removing a single element. The first strategy resulted in a more cumbersome formula than the formula produced by the second strategy. However, the descriptive complexity of the formula produced by this latter strategy is higher.    

\subsection{First Strategy}
The idea is to use binary encodings to represent
each node in the graph, and then to compare the binary encodings of two
connected nodes to identify whether they differ exactly in $1$ bit.
Following a top down approach to the problem, we start with a very general
schema of the formula and then we explore the main sub-formulae involved
in the solution. We aim for a good balance between
level of detail and clarity of presentation. Consequently, 
we leave out of the presentation some trivial sub-formulae which are not central to the general strategy.

Let $\bf G$ be an undirected graph with $|V| = n$. The following second-order formula is satisfied by $\bf
G$ iff $\bf G$ is an \emph{$m$-hypercube graph} for some $m$.
\begin{align*}
\varphi_1 \equiv \exists \leq \big(\mathrm{A1} \wedge \exists F \, \exists m \big(\mathrm{A2} \wedge \forall x y (E(x, y) \leftrightarrow \mathrm{A3}) \wedge \mathrm{A4}\big)\big) \text{ where}
\end{align*} 
\begin{itemize}[leftmargin=*]
\item A1 expresses ``$\leq$ is a total order of the domain $V$ of $\bf G$''.
\item A2 expresses ``$F$ is a bijection on $V$''.
\item A3 expresses ``The binary encodings of $F(x)$ and $F(y)$ have both length $m$ and differ exactly in one bit''.
\item A4 expresses ``There is a node whose binary encoding contains no zeros''.
\end{itemize}
The total order $\leq$ is used to identify each individual node of $V$. Thus,
we can assume that $V = \{0, \ldots, n-1\}$. This is needed for the
binary encoding of the nodes in $V$, as it will become clear latter on.   
It should be clear how to express~A1 and~A2 in the language of second-order
logic. Thus we concentrate our effort in explaining the strategy to express~A3. 
Finally, note that A4 means that all binary encodings (of length $m$) correspond to some node in $V$, which implies that the number of nodes of ${\bf G}$ is a power of $2$, and also that $m = \log_2 n$. A sub-formula that expresses A4 can be easily built by using the same ideas that we use for A3 below. That is, we can existentially quantify for some node $z$, a linear digraph $(V_z, E_z)$ and a Boolean assignment $B_z$ which assigns $1$ to each node, and such that the binary string represented by $(V_z, E_z, B_z)$ is the binary encoding of $F(z)$.

The following formula expresses A3.
\begin{alignat*}{3}
\exists V_x E_x V_y E_y B_x B_y \big(&\mathrm{A3.1}&& \wedge \mathrm{A3.2} \wedge \mathrm{A3.3} \wedge \mathrm{A3.4} \wedge \mathrm{A3.5} \wedge\\
&\exists G \big(&&\mathrm{A3.6} \wedge \\
& &&\forall u v ((E_x(u, v) \rightarrow \exists u' v' (G(u, u') \wedge G(v, v') \wedge E_y(u', v'))) \wedge\\
& && \quad\quad (E_y(u, v) \rightarrow \exists u' v' (G(u', u) \wedge G(v', v) \wedge E_x(u', v')))) \wedge\\
& && \exists v \forall v' ((\mathrm{A3.7} \rightarrow v' \neq v) \wedge\\
& && \quad\quad\quad (\mathrm{A3.8} \rightarrow v' = v))\big)\big) \text{ where}
\end{alignat*}
\begin{itemize}[leftmargin=*]
\item A3.1 expresses ``$(V_x, E_x)$ and $(V_y, E_y)$ are linear digraphs''.
\item A3.2 expresses ``$B_x$ is a function from $V_x$ to $\{0,1\}$''.
\item A3.3 expresses ``$B_y$ is a function from $V_y$ to $\{0,1\}$''.
\item A3.4 expresses ``$(V_x, E_x, B_x)$ is the binary encoding of $F(x)$''. 
\item A3.5 expresses ``$(V_y, E_y, B_y)$ is the binary encoding of $F(y)$''.
\item A3.6 expresses ``$G$ is a bijection from $V_x$ to $V_y$''.
\item A3.7 expresses ``$B_x(v') = B_y(G(v'))$''.
\item A3.8 expresses ``$B_x(v') \neq B_y(G(v'))$''.
\end{itemize}
\begin{figure}\label{encoding}
\psfrag{x1}{\scriptsize$x_1$}
\psfrag{x2}{\scriptsize$x_2$}
\psfrag{xm}{\scriptsize$x_m$}
\psfrag{W}{\scriptsize$W_x$}
\psfrag{b1}{\scriptsize$b_1=0$}
\psfrag{b2}{\scriptsize$b_2=0$}
\psfrag{bm}{\scriptsize$b_m=0$}
\psfrag{b11}{\scriptsize$b_1=1$}
\psfrag{b21}{\scriptsize$b_2=1$}
\psfrag{bm1}{\scriptsize$b_m=1$}
\psfrag{0}{\scriptsize$0$}
\psfrag{20}{\scriptsize$2^0$}
\psfrag{2m2}{\scriptsize$2^{m-2}$}
\psfrag{2m1}{\scriptsize$2^{m-1}$}
\psfrag{VE}{\scriptsize$(V', E')$}
\centering\epsfig{file=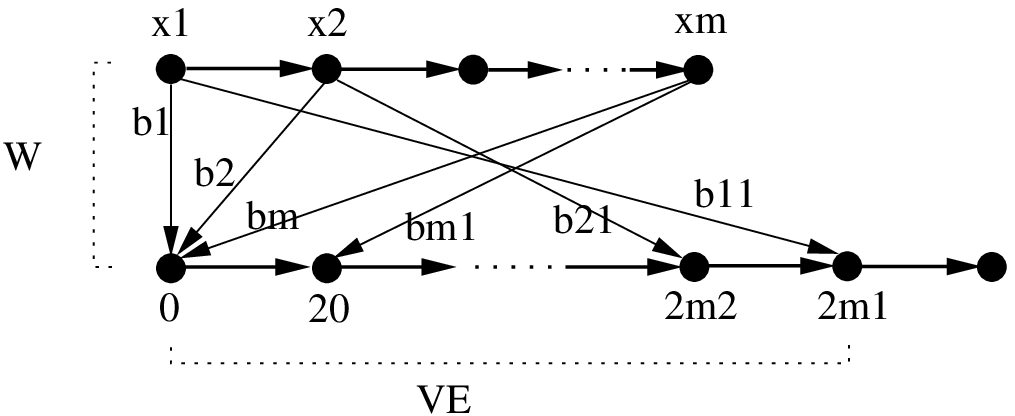,height=4cm}
\begin{center}      
 Figure~4.1  
\end{center}
\end{figure}
To complete the picture, we need to explain how to write~A3.4 and~A3.5 in the language of second-order logic. 
Since both can be expressed in second-order logic in a similar way, we only show the formula for~A3.4. Let $x_i$ be the $i$-th node in the
linear graph $(V_x, E_x)$ defined in the previous formula. We say that $(V_x, E_x, B_x)$ is the binary
encoding of $F(x)$ if 
\[F(x) = b_1 \times 2^{m-1} + b_2 \times 2^{m-2} + \cdots + b_m \times 2^0
\text{, where $b_i = B_x(x_i)$.}\]
In second-order logic, we use a 
function $W_x$ which
assigns to each node $x_i$ in $V_x$ its corresponding value $b_i \times
2^{m-i}$ in the encoding. This function is depicted in
Figure~4.1. The following formula defines the encoding.\\[0.2cm]
$\exists W_x \, I_x \, n_x \, v \, w \, \forall x' \big( \mathrm{A3.4.1} \wedge$\\[0.1cm]
\hspace*{0.3cm} $\forall s \, s' \, q \, q' ((E_x(s, q) \wedge I_x(s, s')
\wedge I_x(q, q')) \rightarrow \mathrm{succ}(s', q')) \wedge$\\[0.1cm] 
\hspace*{0.3cm} $\mathrm{A3.4.2} \wedge \mathrm{A3.4.3} \wedge \, I_x(w, m) \wedge$\\[0.1cm]
\hspace*{0.3cm} $\exists V' \, E' \, v_1 \, v_2 \, w [ \mathrm{A3.4.4} \wedge \mathrm{A3.4.5} \wedge$\\[0.1cm]
\hspace*{0.6cm} $\forall u \big(\neg V'(u) \vee ((u=v_1 \rightarrow u = 0) \wedge (u = v_2
\rightarrow u = 1) \wedge (u=w \rightarrow \mathrm{A3.4.6}) \wedge$\\[0.1cm]
\hspace*{1.2cm} $((u \neq v_1 \wedge u \neq v_2) \rightarrow \exists y'(E'(y',u) \wedge \mathrm{A3.4.7})))\big) \wedge$\\[0.1cm]  
\hspace*{0.6cm} $V_x(x') \rightarrow [(\mathrm{A3.4.8}) \vee (\mathrm{A3.4.9} \, \wedge \exists t (W_x(x', t) \, \wedge \, \mathrm{A3.4.10}))] \wedge$\\[0.1cm]
\hspace*{0.6cm} $\mathrm{A3.4.11} \wedge \mathrm{A3.4.12}]\big)$ where 
\begin{itemize}[leftmargin=*]
\item A3.4.1 expresses ``$I_x$ is a bijection from $V_x$ to $\{1, \ldots, m\}$''.
\item A3.4.2 expresses ``$v$ and $w$ are the first and last nodes of $(V_x, E_x)$, respectively''.
\item A3.4.3 expresses ``$I_x(v, 1)$''.
\item A3.4.4 expresses ``$(V', E')$ is a linear graph''.
\item A3.4.5 expresses ``$v_1, v_2$ and $w$ are the $1$-st, $2$-nd and last nodes in $(V', E')$, respectively''.
\item A3.4.6 expresses ``$\mathrm{exp}(2,m-1,u)$''.
\item A3.4.7 expresses ``$\mathrm{times}(2,y',u)$''.
\item A3.4.8 expresses ``$B_x(x') = 0$'' $\wedge$ ``$W_x(x') = 0$''.
\item A3.4.9 expresses ``$B_x(x') = 1$'' $\wedge$ ``$\mathrm{sum}(n_x, I_x(x'), m)$''.
\item A3.4.10 expresses ``$\mathrm{exp}(2, n_x, t)$''
\item A3.4.11 expresses ``$F(x) = W_x(x_1) + W_x(x_2) + \cdots + W_x(x_m)$ for $x_i$ the $i$-th node in $(V_x, E_x)$''.
\item A3.4.12 expresses ``$W_x$ is a function from $V_x$ to $V'$''.
\end{itemize}
Finally, we note that A3.4.11 can be expressed as follows.\\[0.2cm]
\noindent
$\exists U_x \big(\mathrm{A3.4.11.1} \wedge \forall x' \big(\neg V_x(x') \vee (\mathrm{A3.4.11.2} \wedge \mathrm{A3.4.11.3} \wedge$ \\ 
\hspace*{3.1cm} $(\mathrm{A3.4.11.4} \rightarrow \exists x''(E_x(x'',x') \wedge \mathrm{A3.4.11.5})))\big)\big)$ where
\begin{itemize}[leftmargin=*]
\item A3.4.11.1 expresses ``$U_x$ is a function from $V_x$ to $V$''.
\item A3.4.11.2 expresses ``if $x'$ is the first node in $(V_x,E_x)$ then $U_x(x') = W_x(x')$''.
\item A3.4.11.3 expresses ``if $x'$ is the last node in $(V_x, E_x)$ then $U_x(x') = F(x)$''.
\item A3.4.11.4 expresses ``$x'$ is not the first node in $(V_x, E_x)$''.
\item A3.4.11.5 expresses ``$\mathrm{sum}(U_x(x''), W_x(x'), U_x(x'))$''.
\end{itemize}

\subsection{Second Strategy}
The second strategy to define the class of hypercube graphs can be described in two steps.
\begin{enumerate}[label=\roman*.]
\item To identify every node $x$ in the input graph $\bf G$ with a different subset $S_x$ of a set $V' \subset V$ of cardinality $\log_2 |V|$, making sure that every subset of $V'$ is assigned to some node of $\bf G$. 
\item To check that for every pair of nodes $x$ and $y$ in $\bf G$,  there is an edge between $x$ and $y$ iff $S_x$ can be obtained from $S_y$ by adding or removing a single element.
\end{enumerate} 
In second-order logic we can express this strategy as follows.
\begin{align*}
\varphi_2 \equiv \exists R \big(& \exists V' \big(\mathrm{A1} \wedge \forall S (\mathrm{A2} \rightarrow (\exists x (\mathrm{A3} \wedge \mathrm{A4}))) \wedge \exists z (\mathrm{A5})\big) \wedge\\
& \forall x y ((E(x,y) \wedge E(y,x)) \leftrightarrow \mathrm{A6})\big)
\end{align*}
where
\begin{itemize}
\item A1 expresses ``$V' \subset V \wedge V' \neq \emptyset$''.  
\item A2 expresses ``$S \subseteq V' \wedge S \neq \emptyset$''.
\item A3 expresses ``$x$ is identified with $S$ via $R$''. \\[0.1cm]
$\mathrm{A3} \equiv \forall v (R(x,v) \leftrightarrow S(v))$
\item A4 expresses ``no other node $y \neq x$ can be identified with $S$ via $R$''. \\[0.1cm]
$\mathrm{A4} \equiv \neg \exists y (x \neq y \wedge \forall v (R(y,v) \leftrightarrow S(v)))$
\item A5 expresses ``all nodes, with the only exception of node $z$, are identified with some nonempty subset of $V'$ via R''.\\[0.1cm]
$\mathrm{A5} \equiv \neg \exists v \big(R(z, v)\big) \wedge \forall z' \big(z \neq z' \rightarrow \exists S (\mathrm{A5.1} \wedge \forall v (R(z', v) \leftrightarrow S(v)))\big)$ where
\begin{itemize}
\item A5.1 expresses ``$S \neq \emptyset \wedge S \subseteq V'$''.
\end{itemize}
\item A6 expresses ``the set $S_x$ identified with $x$ can be obtained from the set $S_y$ identified with $y$ by adding or removing a single element''.\\[0.1cm]
$\mathrm{A6} \equiv \exists v \big(\big((R(x,v) \wedge \neg R(y,v)) \vee (R(y,v) \wedge \neg R(x,v))\big) \wedge$\\[0.1cm]
\hspace*{1.4cm} $\forall v' \big(v' \neq v \rightarrow (R(x,v') \leftrightarrow R(y, v'))\big)$ 
\end{itemize}

\begin{remark}
The formula $\varphi_2$ that expresses the second strategy has a prefix of second-order quantifiers of the form $\exists R \exists V' \forall S$. Thus, it is in the class $\Sigma^1_2$. The existence of a formula in $\Sigma^1_1$ that expresses this second strategy is unlikely, since we must express that \emph{every subset} $S$ is identified with some node in the graph. On the other hand, the formula $\varphi_1$ that expresses the first strategy, while considerably more cumbersome than $\varphi_2$, only uses existential second-order quantifiers and can be translated in a rather straightforward way into an equivalent $\Sigma^1_1$ formula. That is, we could transform the current quantification schema of the form \[\forall x y \big(\exists V^1_x E^2_x B^2_x V^1_y E^2_y B^2_y \ldots \exists W^2_x U^2_x W^2_y U^2_y \ldots \big),\] where the superindices added to the relation variables denote their arity, into an schema of the form \[\big(\exists V^2_x E^3_x B^3_x V^2_y E^3_y B^3_y \ldots \exists W^3_x U^3_x W^3_y U^3_y \ldots\big),\] where the prefix ``$\forall x y$'' is eliminated and the arity of every relation variable is increased in $1$, so that we can incorporate all nodes. Thus, for instance, every set $V^1_x$ corresponding to some node $x$ in a graph ${\bf G}$ is now encoded in the binary relation $V^2_x$ in such a way that $V^1_x = \{y |(x,y) \in V^2_x\}$. Then, we can simply express that the set $\{x | (x, y) \in V^2_x\}$ contains every node in the graph $\bf G$. Moreover, we can now omit $V_y, E_y, B_y, W_y$ and $U_y$, since for every pair of nodes $x$ and $y$, their corresponding sets $V^1_x$ and $V^1_y$ will be both encoded into the binary relation $V^2_x$, and something similar will happen for the relations $E$, $B$, $W$ and $U$. 

This is an important consideration since by Fagin-Stockmeyer characterization of the polynomial-time hierarchy \cite{[Sto76]} $\Sigma^1_1$ captures NP while $\Sigma^1_2$ captures $\mathrm{NP}^\mathrm{NP}$.
\end{remark}

\section{Quantified Boolean Formulae}

A \emph{Boolean variable} is any symbol to which we
can associate the truth values 0 and 1. Let $V$ be a countable set of Boolean
variables. The class of \emph{Boolean formulae} over $V$ is the smallest class which
is defined by:
\begin{itemize}
\item The Boolean constants 0 and 1 are Boolean formulae.
\item Every Boolean variable $x$ in $V$ is a Boolean formula.
\item If $\varphi$ and $\psi$ are Boolean formulae then $(\varphi \wedge
\psi)$, $(\varphi \vee \psi)$ and $\neg (\varphi)$ are Boolean formulae.
\end{itemize}

The semantics of the Boolean formulae is given by the well-known semantics of the propositional logic. 

A \emph{quantified Boolean formula} over $V$, as defined by the influential Garey and
Johnson book on the theory of NP-Completeness \cite{[GJ79]}, is a formula of the form
\[Q_1 x_1 Q_2 x_2 \ldots Q_n x_n (\varphi),\]
where $\varphi$ is a Boolean formula over $V$, $n \geq 0$, $x_1, \ldots, x_n
\in V$ and, for $1 \leq i \leq n$, $Q_i$ is either ``$\exists$'' or ``$\forall$''. 
A variable that occurs in the Boolean formula but does not occur in the
prefix of quantifiers is called a \emph{free variable}. We call QBF the set of
quantified Boolean formulae without free variables. As usual, for $k \geq 1$, $\mathrm{QBF}_k$ denotes
the fragment of QBF which consists of those formulae which start with an
existential block and have $k$ alternating blocks of quantifiers. Let $X
\subset V$ be a
finite set of Boolean variables, we assume w.l.o.g. that a formula in
$\mathrm{QBF}_k$ over $X$ is of the form \[\exists \bar{x}_1 \forall \bar{x}_2 \ldots
Q \bar{x}_k (\varphi),\]
where for $1 \leq i \leq k$, $\bar{x}_i = (x_{i1}, \ldots, x_{il_i})$ is a
vector of $l_i$ different variables from $X$, $\exists \bar{x}_i$ denotes a
block of $l_i$ quantifiers of the form $\exists x_{i1}, \ldots, \exists x_{il_i}$, $\forall \bar{x}_i$ denotes a
block of $l_i$ quantifiers of the form $\forall x_{i1}, \ldots, \forall x_{il_i}$, $\varphi$ is a (quantifier free) Boolean formula 
over $X$, $Q$ is ``$\exists$'' if $k$ is odd and ``$\forall$'' if $k$ is even, and the sets $X_1, \ldots, X_k$ of variables 
in $\bar{x}_1, \ldots, \bar{x}_k$, respectively, form a partition of $X$.    

We define next the notion of satisfiability of quantified Boolean
formulae. But first we introduce the concept of alternating valuations which
uses rooted binary trees to represent all possible valuations for a given formula, and paths 
from the root to the leaves of such trees to represent individual valuations. 
This unusual way of representing valuations is motivated by the way in
which we express in second-order logic the satisfiability problem for the
classes $\mathrm{QBF}_k$.    

Let $\mathbf{T}_v$ be a rooted binary tree of
vocabulary $\sigma_{\mathbf{T}_v} = \{E, B, 0, 1\}$. That is, $\mathbf{T}_v$
is a maximally connected acyclic digraph in which every vertex has at most two child
vertices and, except for the root, has a unique parent. Here, $0$ and $1$ are
constant symbols which are interpreted as truth values and $B^{\mathbf{T}_v}$
is a total function which assigns a truth value $0^{\mathbf{T}_v}$ or
$1^{\mathbf{T}_v}$ to each vertex in $V$. 
We say that $\mathbf{T}_v$ is an \emph{alternating valuation} if the
following holds:
\begin{itemize}
\item Every leaf of $\mathbf{T}_v$ is at the same depth $d$. 
\item All vertices at a given depth, i.e., in the same level, have the same
out-degree.  
\item If two vertices $a, b \in V$ are siblings, then $B^{\mathbf{T}_v}(a)
\neq B^{\mathbf{T}_v}(b)$.
\end{itemize}

Let $\varphi \equiv \exists \bar{x}_1 \forall \bar{x}_2 \ldots Q \bar{x}_k
(\psi)$ be a formula in $\mathrm{QBF}_k$, where $Q$ is ``$\exists$'' if $k$ is odd and 
``$\forall$'' if $k$ is even, and let $l_j$ for $1 \leq j \leq k$ be the 
length of the $j$-th alternating block of quantifiers. We say that an
alternating valuation $\mathrm{T}_v$ is \emph{applicable} to $\varphi$, if the
depth of $\mathrm{T}_v$ is $l_1 + \cdots + l_k - 1$ and for every $1 \leq i
\leq l_1 + \cdots + l_k$, it holds that:
\begin{itemize}
\item All vertices at depth $i-1$ have no siblings if $1 \leq i \leq l_1$ or
$l_1 + l_2 + 1 \leq i \leq l_1 + l_2 + l_3$ or $\cdots$ or $l_1 + l_2 + \cdots +
l_{k'-1} + 1 \leq i \leq l_1 + l_2 + \cdots + l_{k'}$, where $k' = k$ if the $k$-th
block of quantifiers is existential and $k' = k - 1$ otherwise. 
\item  All vertices at depth $i-1$ have exactly one sibling if $l_1 + 1 \leq i
\leq l_1 + l_2$ or $l_1 + l_2 + l_3 + 1 \leq i \leq l_1 + l_2 + l_3 + l_4$ or
$\cdots$ or $l_1 + l_2 + \cdots +
l_{k''-1} + 1 \leq i \leq l_1 + l_2 + \cdots + l_{k''}$, where $k'' = k$ if the $k$-th
block of quantifiers is universal and $k'' = k - 1$ otherwise. 
\end{itemize}

Let $\gamma = \exists \bar{x}_1 \forall \bar{x}_2 \ldots
Q \bar{x}_k (\varphi)$ be a formula in $\mathrm{QBF}_k$ over $X$, and let $\mathbf{T}_v$ be an alternating
valuation applicable to $\gamma$. A \emph{leaf valuation} $\mathbf{L}_v$ is a linear
subgraph of $\mathbf{T}_v$ of vocabulary $\sigma_{\mathbf{T}_v}$ which corresponds 
to a path from the root to a leaf in $\mathbf{T}_v$. Let $v$ be a mapping from
the set of variables $X$ to $\{0, 1\}$, i.e., a Boolean assignment, such that
for $x_i \in X$ the $i$-th variable in the prefix of quantifiers
of $\gamma$, it holds that $v(x_i) = 1$ iff $B^{\mathbf{L}_v}(n_i) =
1^{\mathbf{L}_v}$ for $n_i$ the $i$-th node in the linear
order induced by $E^{\mathbf{L}_v}$. We say that $\mathbf{L}_v$ \emph{satisfies}
$\gamma$, 
written $\mathbf{L}_v \models \gamma$, if the Boolean assignment $v$
satisfies $\varphi$. That is, if $\varphi$ is a Boolean variable $x_i$ in $X$, then $\mathbf{L}_v \models
\varphi$ if $v(x_i) = 1$; if $\varphi = \neg(\psi)$, then $\mathbf{L}_v
\models \varphi$ if $\mathbf{L}_v \not\models \psi$ (i.e., if it is not the
case that $\mathbf{L}_v \models \psi$); if $\varphi = (\psi \vee \alpha)$, then
$\mathbf{L}_v \models \varphi$ if either $\mathbf{L}_v \models \psi$ or
$\mathbf{L}_v \models \alpha$; and if $\varphi = (\psi \wedge \alpha)$, then
$\mathbf{L}_v \models \varphi$ if both $\mathbf{L}_v \models \psi$ and
$\mathbf{L}_v \models \alpha$.
Finally, we say that the alternating valuation $\mathbf{T}_v$ \emph{satisfies}
$\gamma$ if every leaf valuation $\mathbf{L}_v$ of $\mathbf{T}_v$ satisfies
$\gamma$.   

A Boolean formulae $\varphi$ in $\mathrm{QBF}_k$ is \emph{satisfiable} if and only if
there is an alternating valuation $\mathbf{T}_v$ which satisfies $\varphi$;
otherwise $\varphi$ is \emph{unsatisfiable}. $\mathrm{SATQBF}_k$ is the set of
$\mathrm{QBF}_k$ formulae that are satisfiable. $\mathrm{SATQBF} =
\bigcup_{k \geq 1} \mathrm{SATQBF}_k$.  

It is well known that $\mathrm{SATQBF}_k$ is complete for the level
$\Sigma^p_k$ of the polynomial-time hierarchy (see \cite{[GJ79],[BDG95]} among
others sources). It is also well known that second-order logic captures the
polynomial-time hierarchy. In fact, there is an exact correspondence between
the prenex fragments of second-order logic with up to $k$ alternations of
quantifiers $\Sigma^1_k$ and the levels $\Sigma^p_k$ of the polynomial time
hierarchy \cite{[Sto76]}. Thus, for every $k$, $\mathrm{SATQBF}_k$ can be
defined in second-order logic, in fact, it can even be defined in
$\Sigma^1_k$. Regarding $\mathrm{SATQBF}$, we note that it is
$\mathrm{PSPACE}$-complete \cite{[Sto76]}. Since existential third-order logic
captures $\mathrm{NTIME}(2^{n^{{\cal O}(1)}})$ (see \cite{[HT06]}) and
$\mathrm{PSPACE} \subseteq \mathrm{DTIME}(2^{n^{{\cal O}(1)}}) \subseteq
\mathrm{NTIME}(2^{n^{{\cal O}(1)}})$, we know that $\mathrm{SATQBF}$ can be
defined in existential third-order logic. In the following sections we present
a second-order formula that defines $\mathrm{SATQBF_k}$ and a 
third-order formula that defines $\mathrm{SATQBF}$, respectively. 

\section{$\mathrm{SATQBF_k}$ in Second-Order Logic}\label{SATQBFk}

Following a top-down approach, we present a detailed construction of a second-order formula
that defines $\mathrm{SATQBF}_k$. But first, we need to fix an encoding of
quantified Boolean formulae as relational structures. 

There is a well-known correspondence between words and finite
structures. Let $A$ be a finite alphabet and let $\pi(A)$ be
the vocabulary $\{\leq\} \cup \{R_a : a \in A\}$, where $\leq$ is a binary
relation symbol and the $R_a$ are unary relation symbols. We can identify any word $v = a_1
\ldots a_{n}$ in $A^*$ with a $\pi(A)$-structure
${\bf B}$, where the cardinality of $B$ equals the length of $v$,
$\leq^{\bf B}$ is a total order on ${\bf B}$, and, for each $R_a \in \pi(A)$,
$R_a^{\bf B}$ contains the positions in $v$ carrying an $a$, \\[0.1cm]
\hspace*{1cm}$R_a^{\bf B} = \{b \in B : \textrm{for some $j$ $(1 \leq j \leq
n)$,}$\\[0.1cm]
\hspace*{3.4cm}$\textrm{$b$ is the $j$-th element in the order $\leq^{\bf B}$ and $a_j = a$}
\}$\\[0.1cm]
Such structures are usually known as \emph{word models} for $v$ (\cite{[EF99]}). 
As any two word models for $v$ are isomorphic, we can speak of \emph{the}
word model for $v$. 

Note that we can represent Boolean variables of the form $x_n$ by using a
symbol ``$X$'' followed
by a sequence of $n$ symbols ``$|$''. For instance, we can write $X|||$ for
$x_3$. Thus using \emph{word models}, 
every quantified Boolean formula $\varphi$ can be viewed as a finite
relational structure $G_{\varphi}$ of the
following vocabulary. 
\[\pi = \{\leq, P_\neg, P_\vee, P_\wedge, P_\exists, P_\forall,
P_(, P_), P_X, P_|\}\]

\begin{example}\label{example1}
If $\varphi$ is the quantified Boolean formula $\exists x_1 \forall x_2 ((\neg
x_1) \vee x_2)$, which using our notation for the variables
corresponds to $\exists X| \forall X|| ((\neg X|) \vee X||)$, 
then the following {$\pi$-structure} ${\bf G}_\varphi$ (note that ${\bf
G}_\varphi$ is a linear graph)
where $G_\varphi = \{1, \ldots, 18\}$, $\leq^{{\bf G}_\varphi}$ is a total order on ${\bf G}_\varphi$, 
$P^{{\bf G}_\varphi}_\neg = \{10\}$,
$P^{{\bf G}_\varphi}_\vee = \{14\}$, $P^{{\bf G}_\varphi}_\wedge = \emptyset$,
$P^{{\bf G}_\varphi}_\exists
= \{1\}$, $P^{{\bf G}_\varphi}_\forall = \{4\}$, $P^{{\bf G}_\varphi}_( = \{8,
9\}$, $P^{{\bf G}_\varphi}_)
= \{13, 18\}$, $P^{{\bf G}_\varphi}_X = \{2, 5, 11, 15\}$, $P^{{\bf
G}_\varphi}_| = \{3, 6, 7, 12,
16, 17\}$, encodes $\varphi$.
\end{example}

We show next how to build a second-order logic formula $\varphi_{\mathrm{SATQBF}_k}$
such that, given a relational structure ${\bf G}_\varphi$ of vocabulary $\pi$, it
holds that ${\bf G}_\varphi \models \varphi_{\mathrm{SATQBF}_k}$ iff the
quantified Boolean formula $\varphi$ represented by ${\bf G}_\varphi$, is
satisfiable. That is, we show next how to build a second-order formula
$\varphi_{\mathrm{SATQBF}_k}$ of vocabulary $\pi$ that defines 
$\mathrm{SATQBF}_k$. As mentioned earlier, we follow a top-down approach for the
construction of this formula. At the highest level of abstraction,
we can think of $\varphi_{\mathrm{SATQBF}_k}$ as a second-order formula that
expresses the following:
\begin{equation}\label{eq1}
\text{ ``There is an alternating valuation ${\bf T}_v$ applicable to $\varphi$ that satisfies
$\varphi$''.} \tag{A} 
\end{equation}
Recall that an alternating valuation ${\bf T}_v$ satisfies $\varphi$ iff every
leaf valuation ${\bf L}_v$ of ${\bf T}_v$ satisfies the quantifier-free part
$\varphi^\prime$ of $\varphi$. Also recall that every leaf
valuation ${\bf T}_v$ corresponds to a Boolean assignment $v$. Thus, if  
$\varphi = \exists \bar{x}_1 \forall \bar{x}_2 \ldots Q \bar{x}_k
(\varphi^\prime)$, where for $1 \leq i \leq k$, $\bar{x}_i = (x_{i1}, \ldots,
x_{il_i})$, $Q$ is ``$\exists$'' if $k$ is odd and ``$\forall$'' if $k$ is
even, $X_1, \ldots, X_k$ are the set of variables 
in $\bar{x}_1, \ldots, \bar{x}_k$, respectively, and $\varphi^\prime$ is a
(quantifier free) Boolean formulae 
over $X = X_1 \cup \cdots \cup X_k$, then the expression in (\ref{eq1}) can be divided in two parts:
\begin{itemize}[leftmargin=1cm]
\item[AVS1] (Alternating Valuation that Satisfies $\varphi$, Part 1) which expresses\\[0.2cm]
``There is a partial Boolean assignment $v_1$ on $X_1$,\\[0.1cm] 
\hspace*{0.3cm} such that for all partial Boolean assignments $v_2$ on $X_2$,\\[0.1cm] \hspace*{0.45cm} \ldots,\\[0.1cm]\hspace*{0.6cm} there is (or ``for all'' if k is even) a partial Boolean assignment $v_k$ on $X_k$''.
\item[AVS2] (Alternating Valuation that Satisfies $\varphi$, Part 2) which expresses\\[0.2cm]
``The Boolean assignment $v = v_1 \cup v_2 \cup \cdots \cup v_k$ satisfies the (quantifier free) Boolean formula $\varphi^\prime$''.
\end{itemize}

For each partial Boolean assignment $v_i$ ($1 \leq i \leq k$), we use
a second-order variable $V_i$ of arity one and two second-order variables
$E_i$ and $B_i$ of arity two, to store the encoding of each $v_i$ as a linear graph 
$G_i = (V_i, E_i)$ with an associated function $B_i :
V_i \rightarrow \{0,1\}$ (see Figure~6.1). 
\begin{figure}[h!]\label{figure7}
\psfrag{E1}{\scriptsize$\exists$ linear graph $G_1$}
\psfrag{A1}{\scriptsize$\forall$ linear graph $G_2$}
\psfrag{E2}{\scriptsize$\exists$ linear graph $G_3$ \ldots}
\psfrag{Q}{\scriptsize$Q$ linear graph $G_k$}
\psfrag{G1}{\scriptsize$G_1$}
\psfrag{G2}{\scriptsize$G_2$}
\psfrag{G3}{\scriptsize$G_3$}
\psfrag{Gk}{\scriptsize$G_k$}
\psfrag{G1b}{\scriptsize$G_1 = (V_1, E_1)$}
\psfrag{G2b}{\scriptsize$G_2 = (V_2, E_2)$}
\psfrag{Gkb}{\scriptsize$G_k = (V_k, E_k)$}
\centering\epsfig{file=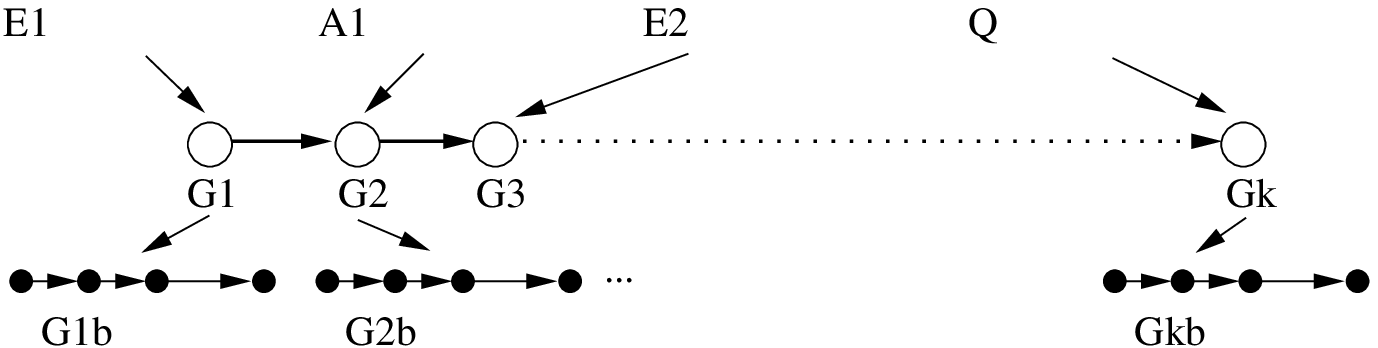,height=3cm}
\begin{center}      
 Figure~6.1
\end{center}
\end{figure}
Correspondingly, we use a second-order variable $V_t$ of arity one and two
second-order variables $E_t$ and $B_t$ of arity two, to store the encoding of
each Boolean assignment $v$ (leaf valuation ${\bf T}_v$) as a linear graph
$G_t = (V_t, E_t)$ with an associated function $B_t :
V_t \rightarrow \{0,1\}$. Figure~6.2 illustrates an alternating
valuation applicable to $\varphi$ and its corresponding encoding.
\begin{figure}[h!]\label{figure8}
\psfrag{formula}{\scriptsize$\exists x_{11} \exists x_{12} \exists x_{13} \cdots \exists x_{1l_1} \forall x_{21} \forall x_{22} \forall x_{23} \cdots \forall x_{2l_2} \exists x_{31} \exists x_{32} \exists x_{33} \cdots \exists x_{3l_3} \cdots Q x_{k1} Q x_{k2} Q x_{k3} \cdots Q x_{kl_k} \big(\varphi'\big)$}
\psfrag{G1}{\scriptsize$G_1 = (V_1, E_1, B_1)$}
\psfrag{G2}{\scriptsize$G_2 = (V_2, E_2, B_2)$}
\psfrag{G3}{\scriptsize$G_3 = (V_3, E_3, B_3) \quad \ldots$}
\psfrag{Gk}{\scriptsize$G_k = (V_k, E_k, B_k)$}
\psfrag{U1}{\scriptsize$U_1$}
\psfrag{U2}{\scriptsize$U_2$}
\psfrag{U3}{\scriptsize$U_3 \quad \ldots$}
\psfrag{Uk}{\scriptsize$U_k$}
\psfrag{Graph}{\scriptsize Graph $G_k$}
\psfrag{01}{\scriptsize$0/1$}
\epsfig{file=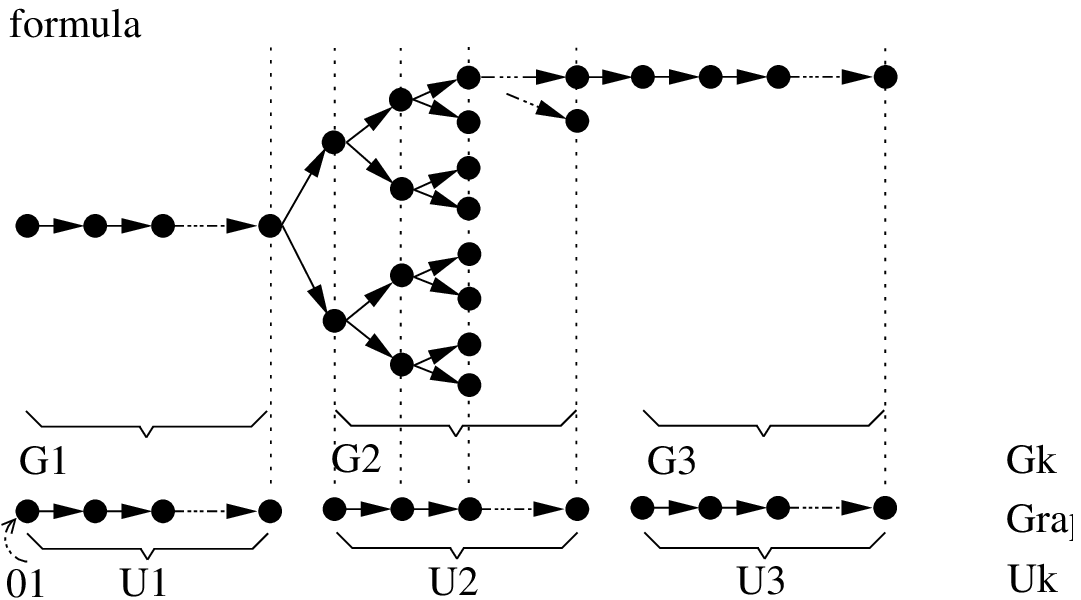,width=10.5cm}
\begin{center}      
 Figure~6.2  
\end{center}
\end{figure}

In the next subsection we describe the process followed to build
a second-order formula to express Statement~AVS1. Then we describe in
Subsection~\ref{expresing22}, the corresponding process for Statement~AVS2.

\subsection{Expressing Statement~AVS1}
Let $k_\exists$ and $k_\forall$ be the index of the last existential
quantifier block and
the last universal quantifier block, respectively, in the prefix of $k$ blocks
of quantifiers of $\varphi$. We can express Statement~AVS1 as follows:\\[0.3cm]
$\exists V_1 E_1 B_1 \forall V_2 E_2 B_2 \cdots Q_k V_k E_k B_k \exists V_t
E_t B_t U_1, U_2, \ldots, U_k \Big( \mathrm{A1} \wedge \mathrm{A2} \wedge \mathrm{A3} \wedge \mathrm{A4} \wedge \mathrm{A5} \wedge$
\begin{flushright} $\big( \big(\mathrm{A6} \wedge \mathrm{A7} \wedge \mathrm{A8} \wedge \mathrm{A9} \wedge  \mathrm{A10} \wedge \mathrm{A11} \big) \rightarrow \mathrm{AVS2} \big)\Big)$ where
\end{flushright}
\begin{itemize}[leftmargin=*]
\item A1 expresses ``$G_t = (V_t, E_t)$ is a linear graph''.
\item A2 expresses ``The length of $G_t$ equals the number of variables that appear in the prefix of quantifiers of $\varphi$''.
\item A3 expresses ``$G_1 = (V_1, E_1), G_3 = (V_2, E_2), \ldots, G_{k_\exists} = (V_{k_\exists}, L_{k_\exists})$ are linear graphs''.
\item A4 expresses ``$B_1: V_1 \rightarrow \{0,1\}$, $B_3: V_3 \rightarrow \{0,1\}$, \ldots, $B_{k_\exists}: V_{k_\exists} \rightarrow \{0,1\}$ are total functions''.
\item A5 expresses ``The lengths of the linear graphs $G_1, G_3, \ldots, G_{k_\exists}$ equal the lengths of their corresponding blocks of quantifiers in $\varphi$''.
\item A6 expresses ``$V_1, V_2, \ldots, V_k$ are pairwise disjoint sets''.
\item A7 expresses ``$G_2 = (V_2, E_2), G_4 = (V_4, E_4), \ldots, G_{k_\forall} = (V_{k_\forall}, L_{k_\forall})$ are linear graphs''
\item A8 expresses ``$B_2: V_2 \rightarrow \{0,1\}$, $B_4: V_4 \rightarrow \{0,1\}$, \ldots, $B_{k_\forall}: V_{k_\forall} \rightarrow \{0,1\}$ are total functions'' 
\item A9 expresses ``The lengths of the linear graphs $G_2, G_4, \ldots, G_{k_\forall}$ equal the lengths of their corresponding blocks of quantifiers in $\varphi$''.
\item A10 expresses ``$U_1$ is a total injection from $G_1$ to the first part of $G_t$ and $U_2$ is a total injection from $G_2$ to the second part of $G_t$ \ldots and $U_k$ is a total injection from $G_k$ to the $k$-th part of $G_t$''.
\item A11 expresses ``$B_t: V_t \rightarrow \{0,1\}$ is a total function that \emph{coincides} with $B_1$, $B_2$, $\ldots$, $B_k$''. 
\item AVS2 expresses Statement~AVS2 as described in Subsection~\ref{expresing22}.
\end{itemize}

Next, we discuss how to write the sub-formulae A1--A11 in second-order logic. 

\begin{enumerate}[label={A}\arabic*., leftmargin=*]

\item This is expressed by the auxiliary formula $\mathrm{LINEAR}(V_t, E_t)$, which is defined
in Subsection~\ref{auxiliary} below.

\item This is implied by the following statement which is expressed in further
detail in Subsection~\ref{exparti}~(A).\\
``There is a partial surjective injection $V_p$ from the quantifier prefix of
$\varphi$ to $G_t$, which maps every $X$ in the prefix to its
corresponding node in $G_t$, and which preserves $\leq^{{\bf G}_\varphi}$ and
$E_t$''. 

\item $\mathrm{LINEAR}(V_1, E_1) \wedge \mathrm{LINEAR}(V_3, E_3) \wedge
\cdots \wedge \mathrm{LINEAR}(V_{k_\exists}, E_{k_\exists})$, where the sub-formulae $\mathrm{LINEAR}(V_i, E_i)$
are as defined in Subsection~\ref{auxiliary}.  

\item $\forall t, p, p^\prime \Big(\bigwedge_{i = 1, 3, \ldots, k_\exists} \big( \mathrm{A4.1} \wedge \mathrm{A4.2} \wedge \mathrm{A4.3} \big) \Big)$
\begin{itemize}
\item A4.1 expresses ``$B_i$ is a function''. \\
$\mathrm{A4.1} \equiv ((B_i(t,p) \wedge B_i(t, p^\prime)) \rightarrow p = p^\prime)$
\item A4.2 expresses  ``$B_i$ is total''.\\
$\mathrm{A4.2} \equiv (V_i(t) \rightarrow \exists p (B_i(t, p)))$
\item A4.3 expresses ``the range of $B_i$ is $\{0, 1\}$''.\\
$\mathrm{A4.3} \equiv (B_i(t, p) \rightarrow (\text{$p = 1$ $\vee$ $p = 0$}))$\\
where $p = 0$ and $p = 1$ have the obvious meaning and are defined in
Subsection~\ref{auxiliary}.
\end{itemize}

\item\label{v} If $k_\exists \neq k$, then \\[0.2cm]
$\bigwedge_{1, 3, \ldots, k_\exists}\big(\exists L' v_1 v_2 \ldots v_{k_\exists}
v_{k_\exists+1} (\alpha_{k_\exists} \wedge \zeta_i)\big)$\\[0.2cm]
where $\alpha_{k_\exists}$ is the formula template $\alpha_i$ instantiated
with $i = k_\exists$.\\[0.2cm]
If $k_\exists = k$, then \\[0.2cm]
$\big(\bigwedge_{1, 3, \ldots, k_\exists-2} \big(\exists L' v_1 v_2 \ldots
v_{k_\exists-1} (\alpha_{k_{\exists-2}} \wedge \zeta_i)\big) \big) \wedge
\exists L' v_1 v_2 \ldots v_k v_e (\beta_1 \wedge \beta_2 \wedge \beta_3)$\\[0.2cm]
where $\alpha_{k_\exists-2}$ is the formula template $\alpha_i$ instantiated
with $i = k_\exists-2$ (Note that $k_\exists - 2$ is the previous to the last
existential block, and the subformulae $\beta_1, \beta_2$ and $\beta_3$ take
care of the last block of quantifiers).\\[0.2cm]
Next, we define the subformulae $\alpha_i$, $\beta_1$ $\zeta_i$, $\beta_2$ and $\beta_3$ in the listed order. For their definitions we use an auxiliary formula $\mathrm{PATH}_\leq(x,y)$ which is in turn defined in Subsection~\ref{auxiliary} below, and which expresses ``the pair $(x,y)$ is in the transitive closure of the relation $\leq$''.\\[0.2cm] 
The subformula $\alpha_i$ is satisfied if, for $1 \leq j \leq i$, $v_j$ is the position of the first quantifier of
the $j$-th block (when $i$ is not the last block of quantifiers).\\[0.2cm] 
$\alpha_i \equiv \big(P_\exists(v_1) \wedge P_\forall(v_2) \wedge \ldots \wedge
P_Q(v_{i+1}) \wedge \neg \exists x (x \neq v_1 \wedge x \leq v_1) \wedge$\\[0.2cm]
\hspace*{0.9cm}$\mathrm{PATH}_\leq(v_1, v_2) \wedge \mathrm{PATH}_\leq(v_2, v_3) \wedge \cdots
\wedge \mathrm{PATH}_\leq(v_{i},v_{i+1}) \wedge$\\[0.2cm]
\hspace*{0.9cm}$\neg \exists x (\mathrm{PATH}_\leq(v_1, x) \wedge \mathrm{PATH}_\leq(x, v_2) \wedge x \neq v_1 \wedge x \neq v_2 \wedge
P_\forall(x)) \wedge$\\[0.2cm]
\hspace*{0.9cm}$\neg \exists x (\mathrm{PATH}_\leq(v_2, x) \wedge \mathrm{PATH}_\leq(x, v_3) \wedge x \neq v_2 \wedge x \neq v_3 \wedge
P_\exists(x)) \wedge$\\[0.2cm]
\hspace*{0.9cm}$\ldots \wedge$\\[0.2cm]
\hspace*{0.9cm}$\neg \exists x (\mathrm{PATH}_\leq(v_{i}, x) \wedge
\mathrm{PATH}_\leq(x, v_{i+1}) \wedge x \neq v_i \wedge x \neq v_{i+1} \wedge P_Q(x))\big)$\\[0.2cm]
where $P_Q$ is $P_\forall$ if $i$ is odd or $P_\exists$ if $i$ is even.\\[0.2cm]
The subformula $\beta_1$ is satisfied if, for $1 \leq j \leq i$, $v_j$ is the position of the first quantifier of
the $j$-th block.\\[0.2cm] 
$\beta_1 \equiv \big(P_\exists(v_1) \wedge P_\forall(v_2) \wedge \ldots \wedge
\mathrm{P1}(v_{k}) \wedge P_|(v_e) \wedge \neg \exists x (x \leq v_1) \wedge$\\[0.2cm]
\hspace*{0.9cm}$\mathrm{PATH}_\leq(v_1, v_2) \wedge \mathrm{PATH}_\leq(v_2, v_3) \wedge \cdots
\wedge \mathrm{PATH}_\leq(v_{k},v_{e}) \wedge$\\[0.2cm]
\hspace*{0.9cm}$\neg \exists x (\mathrm{PATH}_\leq(v_1, x) \wedge \mathrm{PATH}_\leq(x, v_2) \wedge x \neq v_1 \wedge x \neq v_2  \wedge
P_\forall(x)) \wedge$\\[0.2cm]
\hspace*{0.9cm}$\neg \exists x (\mathrm{PATH}_\leq(v_2, x) \wedge \mathrm{PATH}_\leq(x, v_3)  \wedge x \neq v_2 \wedge x \neq v_3 \wedge
P_\exists(x)) \wedge$\\[0.2cm]
\hspace*{0.9cm}$\ldots \wedge$\\[0.2cm]
\hspace*{0.9cm}$\neg \exists x (\mathrm{PATH}_\leq(v_{k}, x) \wedge
\mathrm{PATH}_\leq(x, v_{e})  \wedge x \neq v_k \wedge x \neq v_e  \wedge \mathrm{P2}(x))\big)$\\[0.2cm]
where $\mathrm{P1}$ is $P_\exists$ if $k$ is odd or $P_\forall$ if $k$ is even, and $\mathrm{P2}$ is $P_\forall$ if $k$ is
odd or $P_\exists$ if $k$ is even.\\[0.2cm]
When $i$ is not the index of the last block of quantifiers, the subformula $\zeta_i$ is satisfied if $L^\prime$ is a bijection from the
indices of the symbols $X$ in the $i$-th
alternating block of quantifiers to $V_i$, which preserves $E_i$ and
$\mathrm{Next}_X = \{(a, b) \in \leq^{{\bf G}_\varphi} \, | \, \text{$a$ and $b$
are indices of symbols in the $i$-th block} \wedge  P_X(a) \wedge P_X(b)
\wedge \forall c ((a \leq c \wedge c \leq b) \rightarrow \neg P_X(c))\}$ (i.e., the
order of appearance of the $X$'s in the $i$-th block of quantifiers in the prefix of
$\varphi$). This is illustrated in Figure~6.3. Recall that we encode in $G_i = (V_i, E_i, B_i)$ a partial truth
assignment for the variables in the $i$-th alternating block of
quantifiers.\\[0.2cm]
$\zeta_i \equiv \big(\mathrm{A5.1} \wedge \mathrm{A5.2} \wedge \mathrm{A5.3} \wedge \mathrm{A5.4} \wedge \mathrm{A5.5}\big)$ where
\begin{itemize}
\item A5.1 defines the ``domain of $L^\prime$''.\\
$\mathrm{A5.1} \equiv \forall x \big( (\mathrm{PATH}_\leq(v_i, x) \wedge \mathrm{PATH}_\leq(x, v_{i+1})  \wedge x \neq v_{i+1} \wedge P_X(x)) \leftrightarrow \exists y (L^\prime(x, y)) \big)$
\item A5.2 expresses ``$L^\prime$ is surjective''.\\ 
$\mathrm{A5.2} \equiv \forall y \big(V_i(y) \rightarrow \exists z (L^\prime(z,y))\big)$ 
\item A5.3 expresses ``$L'$ preserves $\mathrm{Next}_X$ and $E_i$'' which implies injectivity.\\
$\mathrm{A5.3} \equiv \forall s t s' t' \Big( \big( L'(s, t) \wedge L'(s',t') \wedge s \neq s'
\wedge \mathrm{PATH}_\leq(v_i, s) \wedge \mathrm{PATH}_\leq(s', v_{i+1})$\\
\hspace*{2.3cm} $\wedge \mathrm{PATH}_\leq(s, s') \wedge \neg \exists z
(\mathrm{PATH}_\leq(s, z) \wedge \mathrm{PATH}_\leq(z, s')  \wedge$\\
\hspace*{2.3cm} $z \neq s \wedge z \neq s'  \wedge P_X(z) ) \big) \rightarrow E_i(t, t') \Big)$ 
\item A5.4 defines the ``range of $L^\prime$''.\\
$\mathrm{A5.4} \equiv \forall x y (L'(x,y) \rightarrow V_i(y))$
\item A5.5 expresses ``$L'$ is a function''.\\
$\mathrm{A5.5} \equiv \forall x y z \big( (L'(x,y) \wedge L'(x,z)) \rightarrow y = z\big)$\\
\end{itemize}
The subformula $\beta_2$ is satisfied if $L^\prime$ is a bijection from the
indices of the symbols $X$ in the $k$-th
alternating block of quantifiers to $V_k$, which preserves $E_k$ and
$\mathrm{Next}_X$ (i.e., the
order of appearance of the $X$'s in the $k$-th block of quantifiers in the prefix of
$\varphi$). \\[0.2cm]
$\beta_2 \equiv \big(\mathrm{A5.1'} \wedge \mathrm{A5.2'} \wedge \mathrm{A5.3'} \wedge \mathrm{A5.4'} \wedge \mathrm{A5.5'}\big)$ where
\begin{itemize}
\item $\mathrm{A5.1'}$ defines the ``domain of $L^\prime$''.\\
$\mathrm{A5.1'} \equiv \forall x \big( (\mathrm{PATH}_\leq(v_k, x) \wedge
\mathrm{PATH}_\leq(x, v_{e}) \wedge P_X(x)) \leftrightarrow \exists y (L^\prime(x, y)) \big)$
\item $\mathrm{A5.2'}$ expresses ``$L^\prime$ is surjective''.\\ 
$\mathrm{A5.2'} \equiv \forall y \big(V_k(y) \rightarrow \exists z
(L^\prime(z,y))\big)$ 
\item $\mathrm{A5.3'}$ expresses ``$L'$ preserves $\mathrm{Next}_X$ and $E_k$'' which implies injectivity.\\
$\mathrm{A5.3'} \equiv \forall s t s' t' \Big( \big( L'(s, t) \wedge L'(s',t') \wedge s \neq s'
\wedge \mathrm{PATH}_\leq(v_k, s) \wedge \mathrm{PATH}_\leq(s', v_{e})$\\
\hspace*{2.3cm} $\wedge \mathrm{PATH}_\leq(s, s') \wedge \neg \exists z
(\mathrm{PATH}_\leq(s, z) \wedge \mathrm{PATH}_\leq(z, s') \wedge$\\
\hspace*{2.3cm} $z \neq s \wedge z \neq s' \wedge P_X(z) ) \big) \rightarrow E_k(t, t') \Big)$ 
\item $\mathrm{A5.4'}$ defines the ``range of $L^\prime$''.\\
$\mathrm{A5.4'} \equiv \forall x y (L'(x,y) \rightarrow V_k(y))$
\item $\mathrm{A5.5'}$ expresses ``$L'$ is a function''.\\
$\mathrm{A5.5'} \equiv \forall x y z \big( (L'(x,y) \wedge L'(x,z)) \rightarrow y = z\big)$\\
\end{itemize}
The last subformula $\beta_3$ is satisfied if $v_e$ is the last symbol ``$|$''
in the prefix of quantifiers of $\varphi$. We use $\mathrm{SUC}_{\leq}(x,y)$ to
denote that $x$ is the immediate successor of $y$ in the total order $\leq^{{\bf
G}_\varphi}$. The formula that expresses $\mathrm{SUC}_{\leq}(x,y)$ is defined in Subsection~\ref{auxiliary}.\\[0.2cm]
$\beta_3 \equiv \Big(\forall v' \big(\mathrm{SUC}_\leq(v_{e}, v') \rightarrow \neg
P_|(v')\big) \wedge P_|(v_{e}) \wedge$\\[0.2cm]
\hspace*{0.8cm} $\forall v' \big(\mathrm{PATH}_\leq(v_{e},v') \rightarrow 
(\neg P_\exists(v') \wedge \neg P_\forall(v')) \big) \wedge$\\[0.2cm]
\hspace*{0.8cm} $\exists x y w \forall v' \big(P_X(x)
\wedge P_Q(w) \wedge \mathrm{SUC}_\leq(x,y) \wedge \mathrm{SUC}_\leq(w,x) \wedge
\mathrm{PATH}_\leq(y, v_{e}) \wedge$\\[0.2cm] 
\hspace*{2.2cm} $((\mathrm{PATH}_\leq(v', v_{e}) \wedge \mathrm{PATH}_\leq(y, v')) \rightarrow
P_|(v'))\big)\Big)$\\[0.2cm]
where $P_Q$ is $P_\exists$ if $k$ is odd, or $P_\forall$ if $k$ is even.\\
\begin{figure}[h!]\label{figure9}
\psfrag{vi}{\scriptsize$v_i$}
\psfrag{s}{\scriptsize$s$}
\psfrag{sprime}{\scriptsize$s'$}
\psfrag{vi1}{\scriptsize$v_{i+1}$}
\psfrag{vk}{\scriptsize$v_k$}
\psfrag{Q}{\scriptsize$Q$}
\psfrag{X}{\scriptsize$X$}
\psfrag{phi}{\scriptsize$\varphi$}
\psfrag{L}{\scriptsize$L' \Rightarrow$}
\psfrag{Next}{\scriptsize ``Next $X$ in the same alternating block'' ($\mathit{Next}_X$)}
\psfrag{t}{\scriptsize$t$}
\psfrag{tprime}{\scriptsize$t'$}
\psfrag{VE}{\scriptsize$(V_i, E_i)$}
\centering\epsfig{file=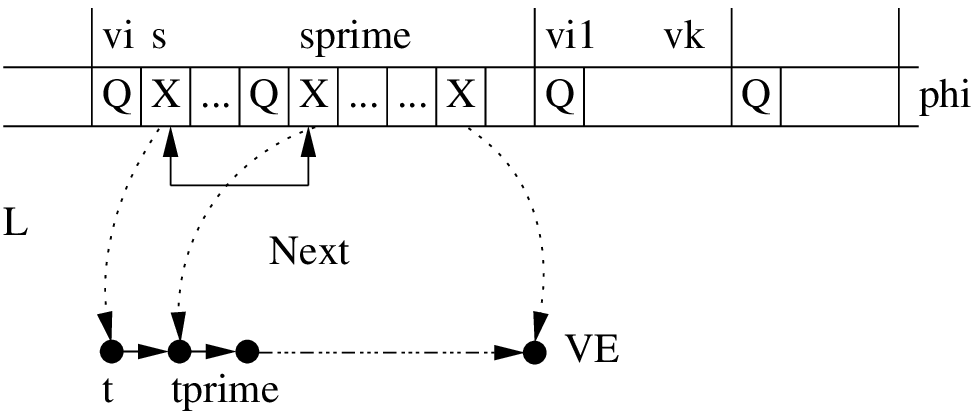,height=4cm}
\begin{center}      
 Figure~6.3  
\end{center}
\end{figure}

\item Let $V_i \cap V_j = \emptyset$ denote $\forall x \big((V_i(x) \rightarrow \neg
V_j(x)) \wedge (V_j(x) \rightarrow \neg V_i(x))\big)$, we can express that $V_1, V_2, \ldots, V_k$ are pairwise disjoint sets as follows.\\[0.2cm]
$(V_1 \cap V_2 = \emptyset) \wedge (V_1 \cap V_3 = \emptyset) \wedge
(V_1 \cap V_4 = \emptyset) \wedge
\cdots \wedge (V_1 \cap V_k = \emptyset) \wedge$\\[0.2cm]
$(V_2 \cap V_3 = \emptyset) \wedge (V_2 \cap V_4 = \emptyset) \wedge
\cdots \wedge (V_2 \cap V_k = \emptyset) \wedge$\\[0.2cm]
$\ldots \wedge (V_{k-1} \cap V_k = \emptyset)$

\item $\mathrm{LINEAR}(V_2, E_2) \wedge \mathrm{LINEAR}(V_4, E_4) \wedge
\cdots \wedge \mathrm{LINEAR}(V_{k_\forall}, E_{k_\forall})$,\\ where  $\mathrm{LINEAR}(V_i, E_i)$
is as defined in Subsection~\ref{auxiliary}.  

\item $\forall t, p, p^\prime \Big( \bigwedge_{i = 2, 4, \ldots, k_\forall} \big( \mathrm{A8.1} \wedge \mathrm{A8.2} \wedge \mathrm{A8.3} \big) \Big)$ where
\begin{itemize}
\item A8.1 expresses ``$B_i$ is a function''.\\
$\mathrm{A8.1} \equiv ((B_i(t,p) \wedge B_i(t, p^\prime) ) \rightarrow p = p^\prime)$
\item A8.2 expresses ``$B_i$ is total''.\\
$\mathrm{A8.2} \equiv (V_i(t) \rightarrow \exists p (B_i(t, p)))$ 
\item A8.3 expresses ``the range of $B_i$ is $\{0, 1\}$''.\\
$\mathrm{A8.3} \equiv (B_i(t, p) \rightarrow (p = 1 \vee p = 0))$\\
where $p = 0$ and $p = 1$ have the obvious meaning and are defined in
Subsection~\ref{auxiliary}.
\end{itemize}

\item If $k_\forall \neq k$, then \\[0.2cm]
$\bigwedge_{2, 4, \ldots, k_\forall}\big(\exists L' v_1 v_2 \ldots
v_{k_\forall} v_{k_\forall+1} (\alpha_{k_\forall} \wedge \zeta_i)\big)$\\[0.2cm]
where $\alpha_{k_\forall}$ is the formula template $\alpha_i$ instantiated
with $i = k_\forall$.\\[0.2cm]
If $k_\forall = k$, then \\[0.2cm]
$\big(\bigwedge_{2, 4, \ldots, k_\forall-2} \big(\exists L' v_1 v_2 \ldots 
v_{k_\forall-1} (\alpha_{k_{\forall-2}} \wedge \zeta_i)\big) \big) \wedge
\exists L' v_1 v_2 \ldots v_k v_e (\beta_1 \wedge \beta_2 \wedge \beta_3)$\\[0.2cm]
where $\alpha_{k_\forall-2}$ is the formula template $\alpha_i$ instantiated
with $i = k_\forall-2$ (Note that $k_\forall - 2$ is the previous to the last
universal block, and the subformulae $\beta_1, \beta_2$ and $\beta_3$ take
care of the last block of quantifiers).\\[0.2cm]
The subformulae $\alpha_i$, $\zeta_i$, $\beta_1$, $\beta_2$ and $\beta_3$ are
the same as in (A5).

\item $\Big(\mathrm{A10.1} \wedge \bigwedge_{2 \leq i \leq k-1} \big( \mathrm{A10.2.i} \big) \wedge \mathrm{A10.3}\Big)$ where
\begin{itemize}
\item A10.1 expresses ``$U_1$ is a total injection from $V_1$ to $V_t$ such that: (a) preserves $E_1$ and $E_t$ and (b) $U_1($``first node in
the order $E_1$''$) = $``first node in the order $E_t$''''.
\item A10.2.i expresses ``$U_i$ is a total injection from $V_i$ to $V_t$ such that: (a) preserves $E_i$ and $E_t$ and (b) $U_i($``first node in the order $E_i$''$) = \mathrm{SUC}_{E_t}(U_{i-1}($``last node in the order $E_{i-1}$''$))$''.
\item A10.3 expresses ``$U_k$ is a total injection from $V_k$ to $V_t$ such that: (a) preserves $E_k$ and $E_t$ and (b) $U_k($``first node in order $E_k$''$) =
\mathrm{SUC}_{E_t}(U_{k-1}($``last node in order $E_{k-1}$''$))$''.
\end{itemize}
We describe next the second-order formula for A10.3 which is in turn illustrated in
Figure~6.4. Note that the node labeled $x$ in Figure~6.4 corresponds to the last node in the linear graph $G_{k-1}$ and that $x$ is mapped by the function $U_{k-1}$ to the node labelled $y$ in the linear graph $G_t$. Accordingly, $U_k$ maps the first node in the linear graph $G_{k}$ (i.e. the node labeled $u$), to the successor of node $y$ in $G_t$ (i.e. to the node labelled $t$). \\[0.2cm]
$\mathrm{A10.3} \equiv \forall x y t u \big(\mathrm{A10.3.1} \wedge \mathrm{A10.3.2} \wedge \mathrm{A10.3.3} \wedge \mathrm{A10.3.4}\big)$ where
\begin{itemize}
\item A10.3.1 expresses ``$U_k$ is a total injection from $V_k$ to $V_t$''.\\
$\mathrm{A10.3.1} \equiv ( (U_k(x,y) \wedge U_k(x,t)) \rightarrow y = t) \wedge$\\
\hspace*{1.5cm} $( (U_k(x,y) \wedge U_k(u, y)) \rightarrow x = u) \wedge$\\
\hspace*{1.5cm} $(V_k(x) \rightarrow \exists y (U_k(x, y))) \wedge$\\
\hspace*{1.5cm} $(U_k(x, y) \rightarrow (V_k(x) \wedge V_t(y)))$
\item A10.3.2 expresses ``preserves $E_t$''.\\
$\mathrm{A10.3.2} \equiv ((U_k(x,y) \wedge U_k(u,t) \wedge E_t(y,t)) \rightarrow E_k(x,u))$
\item A10.3.3 expresses ``preserves $E_k$''.\\
$\mathrm{A10.3.3} \equiv ((U_k(x,y) \wedge U_k(u,t) \wedge E_k(x,u)) \rightarrow E_t(y,t))$
\item A10.3.4 expresses ``$U_k($``first node in order $E_k$''$) = \mathrm{SUC}_{E_t}(U_{k-1}($``last node in order $E_{k-1}$''$))$''.\\
$\mathrm{A10.3.4} \equiv \big( \big(U_{k-1}(x,y) \wedge \neg \exists v (E_{k-1}(x,v)) \wedge E_t(y, t)
\wedge \neg \exists v (E_k(v, u) \wedge V_k(u))\big)$\\ 
\hspace*{1.8cm} $\rightarrow U_k(u,t)\big)$
\end{itemize}
\begin{figure}[h!]\label{figure9bis}
\psfrag{Gk1}{\scriptsize$G_{k-1}$}
\psfrag{Gk}{\scriptsize$G_k$}
\psfrag{x}{\scriptsize$x$}
\psfrag{u}{\scriptsize$u$}
\psfrag{Uk1}{\scriptsize$U_{k-1}$}
\psfrag{Uk}{\scriptsize$U_k$}
\psfrag{y}{\scriptsize$y$}
\psfrag{t}{\scriptsize$t$}
\psfrag{Gt}{\scriptsize$G_t$}
\centering\epsfig{file=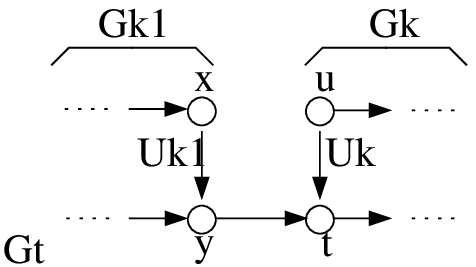,height=4cm}
\begin{center}      
 Figure~6.4 
\end{center}
\end{figure}

\item $\forall x y t p p' \big( (B_1(t,p) \wedge U_1(t,y) \wedge B_t(y, p'))
\rightarrow p = p'\big) \wedge$\\[0.2cm]
$\forall x y t p p' \big( (B_2(t,p) \wedge U_2(t,y) \wedge B_t(y, p'))
\rightarrow p = p'\big) \wedge$\\[0.2cm]
$\ldots \wedge$\\[0.2cm]
$\forall x y t p p' \big( (B_k(t,p) \wedge U_k(t,y) \wedge B_t(y, p'))
\rightarrow p = p'\big)$
\end{enumerate}

\subsection{Expressing Statement~AVS2}\label{expresing22}

Statement~AVS2 can be rephrased as follows:\\[0.2cm]
$\exists \, V_p \, C \, E_C \, \mathit{ST} \, E_\mathit{ST} \, M \, C_\wedge
\, C_\vee \, C_\neg \, C_( \,
C_) \, C_1 \, C_0 \, H_\phi \, \big(\mathrm{AVS2.1} \wedge \mathrm{AVS2.2}\big)$
where
\begin{itemize}
\item AVS2.1 expresses ``There is a Boolean expression $\phi$ which is obtained from the quantifier-free part of $\varphi$ by replacing each occurrence of a variable by the corresponding truth value in $\{0, 1\}$ assigned by the leaf valuation represented by $(G_t, B_t)$''. 
\item AVS2.2 expresses ``The Boolean expression $\phi$ evaluates to true''. 
\end{itemize}

We describe next how to express AVS2.1 and~AVS2.2 in
second-order logic. 

\subsubsection{Expressing AVS2.1}\label{exparti}
The idea is to define mappings to represent the relationships among the input
graph ${\bf G}_\varphi$, the graph $G_t$ and the quantifier-free part of the
input formulae. This is illustrated in Figure~6.5. 
We can express AVS2.1 as follows:\\[0.2cm]
$\mathrm{AVS2.1} \equiv A \wedge B \wedge C$ where 
\begin{figure}\label{figure10}
\psfrag{formula}{\scriptsize$\exists X| \forall X|| \exists X||| \ldots Q X|||\cdots|(\varphi'(X|, X||, X|||, \ldots, X|||\cdots|))$}
\psfrag{V0}{\scriptsize$V_0$: Variable Occurrence}
\psfrag{Input}{\scriptsize Input Graph}
\psfrag{E}{\scriptsize$\exists$}
\psfrag{A}{\scriptsize$\forall$}
\psfrag{X}{\scriptsize$X$}
\psfrag{I}{\scriptsize$|$}
\psfrag{1}{\scriptsize$($}
\psfrag{2}{\scriptsize$)$}
\psfrag{...}{\scriptsize$\ldots$}
\psfrag{Vp}{\scriptsize$V_p$: Variable Position}
\psfrag{Graph}{\scriptsize Graph $G_t$}
\psfrag{Subgraph}{\scriptsize Sub Graph $\phi$}
\psfrag{phi}{\scriptsize$\phi$: A quantifier free formula on: $\{(, ), \wedge, \vee, \neg, 0, 1\}$}
\psfrag{instance}{\scriptsize For instance: $(((1 \wedge (\neg 0)) \vee (1 \wedge 0)) \wedge \ldots (\ldots))$}
\psfrag{H}{\scriptsize$H_\phi$}
\psfrag{01}{\scriptsize$0$/$1$}
\psfrag{w}{\scriptsize$\wedge$}
\psfrag{v}{\scriptsize$\vee$}
\centering\epsfig{file=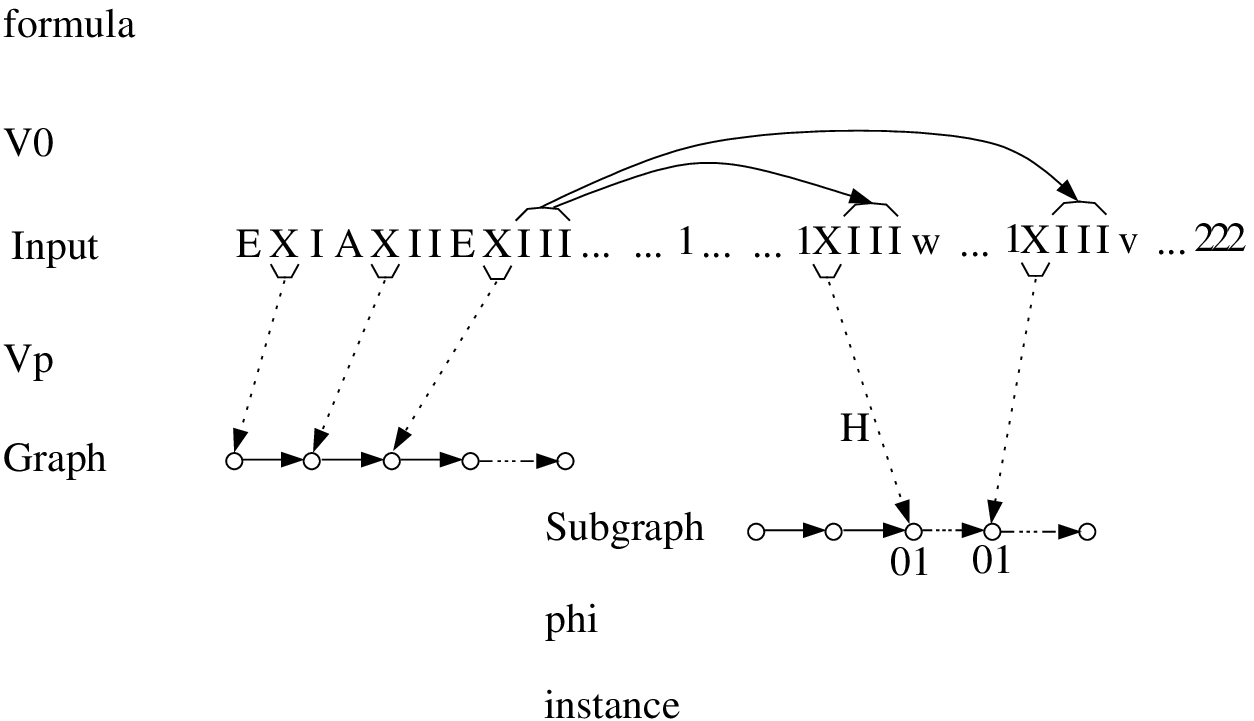,height=6cm}
\begin{center}      
 Figure~6.5  
\end{center}
\end{figure}
\begin{itemize}
\item A expresses ``$V_p$ is a partial bijection from the prefix of quantifiers of
$\varphi$ (restricted to the $X$'s that appear in the prefix) to $V_t$, which
maps every $X$ to its corresponding node in $G_t$, and which preserves  $\leq^{
{\bf G}_\varphi}$ and $E_t$''.\\
$\mathrm{A} \equiv \forall x y z \big(\mathrm{A1} \wedge \mathrm{A2} \wedge \mathrm{A3}\big) \wedge \forall s t s' t' \big(\mathrm{A4}\big)$ where
\begin{itemize}
\item A1 expresses ``$V_p$ is a function''.\\
$\mathrm{A1} \equiv ((V_p(x, y) \wedge V_p(x, z)) \rightarrow y = z)$ 
\item A2 expresses ``$V_p$ is injective''\\
$\mathrm{A2} \equiv ( (V_p(x,y) \wedge V_p(z,y)) \rightarrow x = z)$
\item A3 defines the  ``domain and range of $V_p$''\\ 
$\mathrm{A3} \equiv \big( (P_X(x) \wedge \mathrm{PRED}_\leq(x, z) \wedge
(P_\exists(z) \vee P_\forall(z)) ) \leftrightarrow \exists y (V_t(y) \wedge V_p(x, y)) \big)$   
where $\mathrm{PRED}_\leq(x, z)$ denotes the subformula that expresses that $z$ is the strict predecessor of
$x$ in the order $\leq^{ {\bf G}_\varphi}$ (see Subsection~\ref{auxiliary}). 
\item A4 expresses ``$V_p$ preserves $\leq^{{\bf G}_\varphi}$ and $E_t$''.\\
$(V_p(s,s') \wedge V_p(t, t') \wedge E_t(s',t'))
\rightarrow$\\
\hspace*{0.5cm} $\big(\mathrm{PATH}_\leq(s, t) \wedge \forall z' \big(
(z' \neq s \wedge z' \neq t \wedge \mathrm{PATH}_\leq(s, z') \wedge \mathrm{PATH}_\leq(z', t)) \rightarrow$\\
\hspace*{1cm} $\neg P_X(z')\big) \big)$
\end{itemize}

\item B expresses ``$H_\phi$ is a partial surjective injection from the quantifier
free part of $\varphi$ to the formula $\phi$, encoded as the 
first formula in $(C, E_C)$ (see Figures~6.7 and~6.8),
which maps every $X$ in the quantifier-free part of $\varphi$ to the
corresponding position in the first formula in $(C, E_C)$ (i.e. $\phi$), 
which preserves $\wedge$, $\vee$, $\neg$, $($, $)$, $\leq^{ {\bf G}_\varphi}$ and
$E_C$, and which ignores $|$''.\\[0.1cm]
$\mathrm{B} \equiv \forall x y_1 y_2 z_1 z_2 \big(\mathrm{B1} \wedge \mathrm{B2} \wedge \mathrm{B3} \wedge \mathrm{B4}\big) \wedge \forall x x' z y_1 y_2 z_1 z_2 \big(\mathrm{B5}\big)$ where
\begin{itemize}
\item B1 expresses ``$H_\phi$ is a function''.\\
$\mathrm{B1} \equiv \big( (H_\phi(x,y_1,y_2) \wedge H_\phi(x,
z_1, z_2)) \rightarrow$ \\
\hspace*{2.4cm} $(y_1 = z_1 \wedge y_2 = z_2 \wedge \exists x' (P_{(}(x') \wedge
\mathrm{PATH}_\leq(x',x)) \wedge C(y_1, y_2))\big)$
\item B2 expresses ``$H_\phi$ is injective''. \\
$\mathrm{B2} \equiv \big(H_\phi(x, y_1, y_2) \wedge H_\phi(z, y_1, y_2)
\rightarrow x = z \big)$
\item B3 expresses ``the range of $H_\phi$ is the first formula in $(C, E_C)$''.\\ 
$\mathrm{B3} \equiv \forall y_1' y_2' z_1' z_2' t_1' t_2' v' v_2 \big(\big(\mathit{ST}(v') \wedge \neg \exists y (E_\mathit{ST}(y, v'))
\wedge$ \\
\hspace*{3.8cm} $E_\mathit{ST}(v',v_2) \wedge M(v',y_1',y_2') \wedge M(v_2,
z_1', z_2') \wedge$ \\
\hspace*{3.8cm} $E_C(t_1',t_2',z_1',z_2') \wedge \mathrm{PATH}_{E_C}(y_1',
y_2', y_1, y_2) \wedge $ \\
\hspace*{3.8cm} $\mathrm{PATH}_{E_C}(y_1, y_2, t_1', t_2')\big) \rightarrow
\exists x' (H_\phi(x', y_1, y_2)) \big) \wedge$ 
\item B4 expresses ``the domain of $H_\phi$ corresponds to the quantifier free part of $\varphi$''.
$\mathrm{B4} \equiv \big(\exists x' (P_{(}(x') \wedge \mathrm{PATH}_\leq(x', x)) \rightarrow \exists y_1' y_2' (H_\phi(x, y_1', y_2'))\big)$
\item B5 expresses ``$H_\phi$ preserves $\leq^{ {\bf G}_\varphi}$ (ignoring ``$|$''), $E_C$, $\wedge, \vee$, $(, )$ and $\neg$, and maps X to
$0/1$''. \\
$\mathrm{B5} \equiv \big( (H_\phi(x, y_1, y_2) \wedge
H_\phi(z, z_1, z_2) \wedge E_C(y_1, y_2, z_1, z_2))$\\
\hspace*{1cm} $\rightarrow (\mathrm{SUC_\leq}(x, z) \vee (\mathrm{PATH}_\leq(x,z)
\wedge \forall x' (\mathrm{PATH}_\leq(x, x') \wedge$\\ 
\hspace*{1.6cm}$\mathrm{PATH}_\leq(x', z) \wedge x' \neq x \wedge x' \neq z) \rightarrow
P_|(x')))\big) \wedge$ \\
\hspace*{0.8cm} $\big(H_{\phi}(x, y_1, y_2) \rightarrow ( (P_{(}(x) \wedge
C_{(}(y_1,y_2) ) \vee$ \\
\hspace*{3.6cm} $(P_{)}(x) \wedge C_{)}(y_1,y_2) ) \vee$\\
\hspace*{3.6cm} $(P_{\wedge}(x) \wedge C_{\wedge}(y_1,y_2) ) \vee$\\
\hspace*{3.6cm} $(P_{\vee}(x) \wedge C_{\vee}(y_1,y_2) ) \vee$\\
\hspace*{3.6cm} $(P_{\neg}(x) \wedge C_{\neg}(y_1,y_2) ) \vee$\\
\hspace*{3.6cm} $(P_X(x) \wedge (C_0(y_1,y_2) \vee C_1(y_1,y_2))))\big)$
\end{itemize}

\item C expresses  ``for every bijection $V_0$ from ``$|\cdots|$'' in ``$Q X|\cdots|$'' (where Q is ``$\exists$'' or ``$\forall$'') 
to ``$|\cdots|$'' in ``$(\ldots X|\cdots| \ldots)$''
that links a variable in the quantifier prefix of $\varphi$ with an occurrence of that variable in the quantifier-free part of it, 
the variable in the quantifier free part of $\varphi$ which corresponds to the function $V_0$ is replaced in $\phi$ by the
value assigned to that variable by the leaf valuation $(G_t, B_t)$ (see Figures~6.5)''.
Note that in the formula below, $z_0$ represents the root in
$\mathrm{dom}(V_0)$, $z_f$ represents the leaf in $\mathrm{dom}(V_0)$, $y_0$ represents the root in
$\mathrm{ran}(V_0)$, and $y_f$ represents the leaf in $\mathrm{ran}(V_0)$ (see Figure~6.6). 
Also note that $\phi$ is encoded in $(C, E_C)$ starting in the node $M$(``first node in $(\mathit{ST},
E_\mathit{ST})$'') and ending in the node $E^{-1}_C(M(\text{``second node in $(\mathit{ST},
E_\mathit{ST})$''}))$, and that it is equivalent to the quantifier-free part of $\varphi$
with the variables replaced by $0$ or $1$ according to the leaf valuation
$(G_t, B_t)$ (this is further clarified in Subsection~\ref{part_ii}, also note Figures~6.7 and~6.8). \\[0.1cm]
$\mathrm{C} \equiv \forall\, V_0\, \exists\, z_0 \, y_0 \, z_f \, y_f \, z_0'\, y_0'\, z_f'\, y_f'\, \big( (\mathrm{C1} \wedge \mathrm{C2} \wedge \mathrm{C3} \wedge \mathrm{C4} \wedge \mathrm{C5} \wedge \mathrm{C6}) \rightarrow \mathrm{C7}\big)$
where
\begin{figure}[h!]\label{figure11}
\psfrag{formula}{\scriptsize$\exists X| \forall X|| \exists X||| \ldots Q X|||\cdots|(\varphi'(X|, X||, X|||, \ldots, X|||\cdots|))$}
\psfrag{V0}{\scriptsize$V_0$: Variable Occurrence}
\psfrag{z0}{\scriptsize$z_0'$}
\psfrag{z1}{\scriptsize$z_0$}
\psfrag{z2}{\scriptsize$z_f$}
\psfrag{z3}{\scriptsize$z_f'$}
\psfrag{y0}{\scriptsize$y_0'$}
\psfrag{y1}{\scriptsize$y_0$}
\psfrag{y2}{\scriptsize$y_f$}
\psfrag{y3}{\scriptsize$y_f'$}
\psfrag{X}{\scriptsize$X$}
\psfrag{I}{\scriptsize$|$}
\psfrag{1}{\scriptsize$($}
\psfrag{2}{\scriptsize$)$}
\psfrag{Q}{\scriptsize$Q$}
\psfrag{...}{\scriptsize$\ldots$}
\centering\epsfig{file=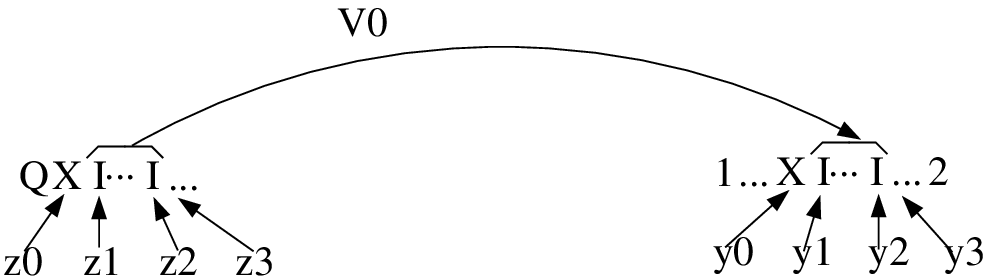,height=3cm}
\begin{center}      
 Figure~6.6  
\end{center}
\end{figure}
\begin{itemize}
\item C1 expresses ``$z_0$ is the root in $\mathrm{dom}(V_0)$, $z_f$ is the leaf in $\mathrm{dom}(V_0)$, $y_0$ is the root in $\mathrm{ran}(V_0)$ and $y_f$ is the leaf in $\mathrm{ran}(V_0)$''.\\
$\mathrm{C1} \equiv V_0(z_0, y_0) \wedge \neg \exists z' y' (\mathrm{PRED}_\leq(z_0, z') \wedge V_0(z', y')) \wedge$\\
\hspace*{0.9cm}$V_0(z_f, y_f) \wedge \neg \exists z' y'
(\mathrm{SUC}_\leq(z_f, z') \wedge V_0(z', y')) \wedge $\\
\hspace*{0.9cm} $\forall z' \big( (\mathrm{PATH}_\leq(z_0, z') \wedge
\mathrm{PATH}_\leq(z', z_f) ) \rightarrow \exists y' (V_0(z', y')) \big) \wedge$\\
\hspace*{0.9cm} $\forall y' \big( (\mathrm{PATH}_\leq(y_0, y') \wedge
\mathrm{PATH}_\leq(y', y_f) ) \rightarrow \exists z' (V_0(z', y')) \big)$
\item C2 expresses ``$V_0$ is a bijection from ``$|\cdots|$'' in ``$Q X|\cdots|$''
to ``$|\cdots|$'' in \\ ``$(\ldots X|\cdots| \ldots)$'' which preserves $\leq^{ {\bf G}_\varphi}$''.\\ 
$\mathrm{C2} \equiv \forall x y v w \big( (V_0(x,y) \rightarrow (P_|(x) \wedge
P_|(y))) \wedge$\\
\hspace*{1.9cm} $( (V_0(x,y) \wedge V_0(x,v)) \rightarrow y = v) \wedge$\\
\hspace*{1.9cm} $( (V_0(x,y) \wedge V_0(w,y)) \rightarrow x = w) \wedge$\\
\hspace*{1.9cm} $( (V_0(x,y) \wedge V_0(v, w) \wedge \mathrm{SUC}_\leq(x,v))
\rightarrow \mathrm{SUC}_\leq(y,w))\big)$
\item C3 expresses ``$z_0'$ is the predecessor of the root in $\mathrm{dom}(V_0)$, i.e., it is the $X$ in the prefix of quantifiers''.\\
$\mathrm{C3} \equiv \mathrm{PRED}_\leq(z_0,z_0') \wedge P_X(z_0')$
\item C4 expresses ``$y_0'$ is the predecessor of the root in $\mathrm{ran}(V_0)$, i.e., it is the $X$ in the quantifier-free part''.\\
$\mathrm{C4} \equiv \mathrm{PRED}_\leq(y_0, y_0') \wedge P_X(y_0')$
\item C5 expresses ``$z_f'$ is the successor of the leaf in $\mathrm{dom}(V_0)$''.\\
$\mathrm{C5} \equiv \mathrm{SUC}_\leq(z_f,z_f') \wedge \neg P_|(z_f')$
\item C6 expresses ``$y_f'$ is the successor of the leaf in $\mathrm{ran}(V_0)$''.\\
$\mathrm{C6} \equiv \mathrm{SUC}_\leq(y_f,y_f') \wedge \neg P_|(y_f')$
\item C7 expresses ``$B_t(V_p(z_0')) = H_\phi(y_0')$''.\\
$\mathrm{C7} \equiv \forall\, x \, x' \,\big( (V_p(z_0',x) \wedge B_t(x, x'))
\rightarrow$ \\
\hspace*{2.8cm} $\exists z_1 z_2 (H_\phi(y_0',z_1,z_2) \wedge $ \\
\hspace*{3.7cm} $( (\text{``$x' = 0$''} \wedge C_0(z_1, z_2)) \vee (\text{``$x'
= 1$''} \wedge C_1(z_1, z_2))))\big)$ 
\end{itemize}
\end{itemize} 

\subsubsection{Expressing AVS2.2}\label{part_ii}

Now we need to check whether the formula $\phi$ built in the previous step,
evaluates to true. The idea is to evaluate one connective at a time, and one pair of
matching parenthesis at a time, until the final result becomes $1$.
Let us look at the example in Figure~6.7. Note that there are ten
evaluation steps, which correspond to ten ``operators'' (i.e., either
connectives or pairs of parenthesis). If there are at most $n$ symbols in
$\phi$, that means that the whole evaluation process needs at most $n$
evaluation steps. This is the reason for using pairs of elements to represent
the nodes of the graph $(C, E_C)$, and quadruples to represent the edges. This
allows the whole evaluation process to take up to $n$ steps (where $n$ is the
length of the input formula). In each step, we have a Boolean sentence on
$\{0, 1\}$ with up to $n$ symbols. Each node in the graph $(\mathit{ST},
E_\mathit{ST})$ represents one such formula, and the function $M$ (for Marker)
is a pointer which tells us in which node in $(C, E_C)$ that formula begins. 
Note that in each evaluation step, either one or
two symbols are removed from the formula at the previous step.   
Figure~6.8 further illustrates the graphs~(A) and~(B) of
Figure~6.7 with a horizontal orientation. Each evaluation step is called a stage. And the first
symbol in each stage is given by the marker function $M$.
\begin{figure}[h!]\label{figure12}
\psfrag{1}{\scriptsize$1.\quad ${\bf (}$((0 \vee 1) \wedge (\neg 0)) \wedge (1 \vee 0))$}
\psfrag{2}{\scriptsize$2.\quad ${\bf (}$((0 \vee 1) \wedge (1)) \wedge (1 \vee 0))$}
\psfrag{3}{\scriptsize$3.\quad ${\bf (}$((0 \vee 1) \wedge 1) \wedge (1 \vee 0))$}
\psfrag{4}{\scriptsize$4.\quad ${\bf (}$((1) \wedge 1) \wedge (1 \vee 0))$}
\psfrag{5}{\scriptsize$5.\quad ${\bf (}$((1) \wedge 1) \wedge (1))$}
\psfrag{6}{\scriptsize$6.\quad ${\bf (}$(1 \wedge 1) \wedge (1))$}
\psfrag{7}{\scriptsize$7.\quad ${\bf (}$(1 \wedge 1) \wedge 1)$}
\psfrag{8}{\scriptsize$8.\quad ${\bf (}$(1) \wedge 1)$}
\psfrag{9}{\scriptsize$9.\quad ${\bf (}$1 \wedge 1)$}
\psfrag{10}{\scriptsize$10.\quad ${\bf (}$1)$}
\psfrag{11}{\scriptsize$11.\quad 1 \, (\mathrm{TRUE})$}
\psfrag{symbols}{\scriptsize$\leq n_{\mathrm{Symbols}}$}
\psfrag{a}{\scriptsize (A)}
\psfrag{steps}{\scriptsize$\leq n_{\mathrm{Steps}}$}
\psfrag{markers}{\scriptsize Markers (in Bold)}
\psfrag{st}{\scriptsize$(\mathit{ST}, E_{\mathit{ST}})$}
\psfrag{includes}{\scriptsize Includes}
\psfrag{evaluation}{\scriptsize An evaluation step}
\psfrag{b}{\scriptsize (B)}
\psfrag{ce}{\scriptsize$(C, E_c)$}
\psfrag{p1}{\scriptsize$(1, 1)$}
\psfrag{p2}{\scriptsize$(1, 2)$}
\psfrag{p3}{\scriptsize$(1, 3)$}
\psfrag{p4}{\scriptsize$(1, n_1)$}
\psfrag{p5}{\scriptsize$(2, 1)$}
\psfrag{p6}{\scriptsize$(2, 2)$}
\psfrag{p7}{\scriptsize$(2, 3)$}
\psfrag{p8}{\scriptsize$(2, n_2)$}
\psfrag{p9}{\scriptsize$(s, 1)$}
\psfrag{p10}{\scriptsize$(s, 2)$}
\psfrag{p11}{\scriptsize$(s, 3)$}
\psfrag{p12}{\scriptsize$(s, n_s)$}
\psfrag{s}{\scriptsize$s \leq n$}
\psfrag{n2}{\scriptsize$\leq n^2$}
\centering\epsfig{file=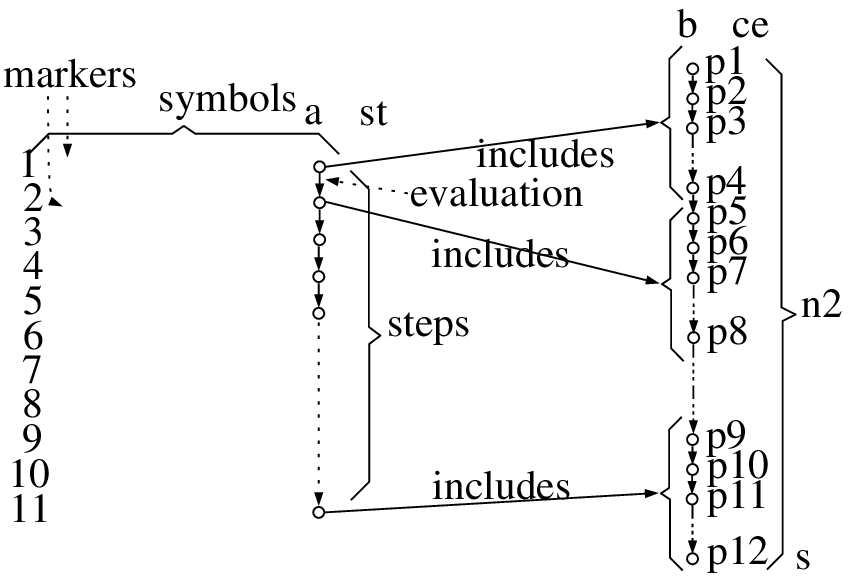,width=12cm}
\begin{center}      
 Figure~6.7 
\end{center}
\end{figure}
\begin{figure}[h!]\label{figure13}
\psfrag{phi}{\scriptsize$\phi$}
\psfrag{E}{\scriptsize$E_v$}
\psfrag{compu}{\scriptsize Computation}
\psfrag{f1}{\scriptsize$(f_1, f_2)$}
\psfrag{f2}{\scriptsize$(l_1, l_2)$} 
\psfrag{f3}{\scriptsize$(f_1', f_2')$} 
\psfrag{f4}{\scriptsize$(l_1', l_2')$} 
\psfrag{ce}{\scriptsize$(C, E_C)$}
\psfrag{n2}{\scriptsize$(\leq n^2)$} 
\psfrag{stages}{\scriptsize Stages}
\psfrag{st}{\scriptsize$(\mathit{ST}, E_{\mathit{ST}})$}
\psfrag{marker}{\scriptsize Marker}
\psfrag{n}{\scriptsize$(\leq n)$} 
\psfrag{M}{\scriptsize$M$}    
\centering\epsfig{file=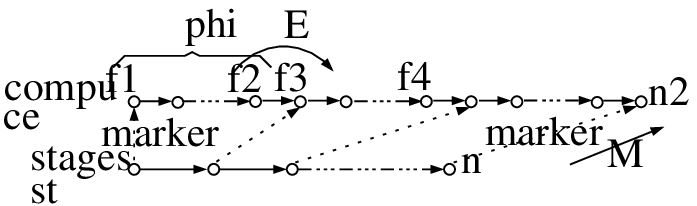,height=3.5cm}
\begin{center}      
 Figure~6.8  
\end{center}
\end{figure}

Based on this description, we can express AVS2.2 in Section~\ref{expresing22} as follows:\\[0.2cm]
$\mathrm{A1} \wedge \mathrm{A2} \wedge \mathrm{A3} \wedge \mathrm{A4} \wedge \mathrm{A5}$ where
\begin{itemize}[leftmargin=*]
\item A1 expresses ``$(C, E_C)$ is a linear graph''.
\item A2 expresses ``$(\mathit{ST}, E_\mathit{ST})$ is a linear graph''.
\item A3 expresses ``$M: \mathit{ST} \rightarrow C$ is an injective and total function that preserves $\mathrm{PATH}$ in $E_{\mathit{ST}}$ and $E_C$''.
\item A4 expresses ``$C_\wedge, C_\vee, C_\neg, C_(, C_), C_0, C_1$ are pairwise disjoint, and $C_\wedge \cup C_\vee \cup C_\neg \cup C_( \cup C_) \cup C_0 \cup C_1 = C$''.
\item A5 expresses ``For every stage $x$, from stage $x$ to stage $x+1$, we need to follow the rules of evaluation (see Figure~6.7 part A). The formula in $(C, E_C)$ at stage $x+1$ is the same as the formula at stage $x$, except for one of three possible sorts of changes, which correspond to the cases (a), (b) and (c) of Figure~6.9''.
\end{itemize} 
\begin{figure}[h!]\label{figure14}
\psfrag{1}{\scriptsize$($} 
\psfrag{b1}{\scriptsize$b_1$}
\psfrag{z}{\scriptsize$\theta$}
\psfrag{b2}{\scriptsize$b_2$}
\psfrag{2}{\scriptsize$)$}
\psfrag{b3}{\scriptsize$b_3$} 
\psfrag{n}{\scriptsize$\neg$}
\psfrag{v1}{\tiny$(v_1, v_2)$}
\psfrag{p1}{\tiny$(p_{11}, p_{12})$} 
\psfrag{p2}{\tiny$(p_{21}, p_{22})$} 
\psfrag{p3}{\tiny$(p_{31}, p_{32})$}
\psfrag{w1}{\tiny$(w_1, w_2)$}  
\psfrag{function}{\scriptsize Function: $E_v$}
\psfrag{where}{\scriptsize where $\theta \in \{\wedge, \vee\}$ and $b_1, b_2, b_3 \in \{0, 1\}$}
\psfrag{A}{\scriptsize {\bf Case (a):} $((p_{21}, p_{22})(p_{31},p_{32})\mathrm{void})$}
\psfrag{B}{\scriptsize {\bf Case (b):} $((p_{21}, p_{22})\mathrm{void})$} 
\psfrag{C}{\scriptsize {\bf Case (c):}}
\psfrag{C2}{\scriptsize$((v_{1}, v_{2})(w_{1},w_{2})\mathrm{void})$}
\psfrag{prima1}{\tiny$(v_1', v_2')(p_{11}', p_{12}')(w_1', w_2')$}
\psfrag{prima2}{\tiny$(v_1', v_2')(p_{11}', p_{12}')(w_1', w_2')$}         
\psfrag{prima3}{\tiny$(p_{11}', p_{12}')$}
\centering\epsfig{file=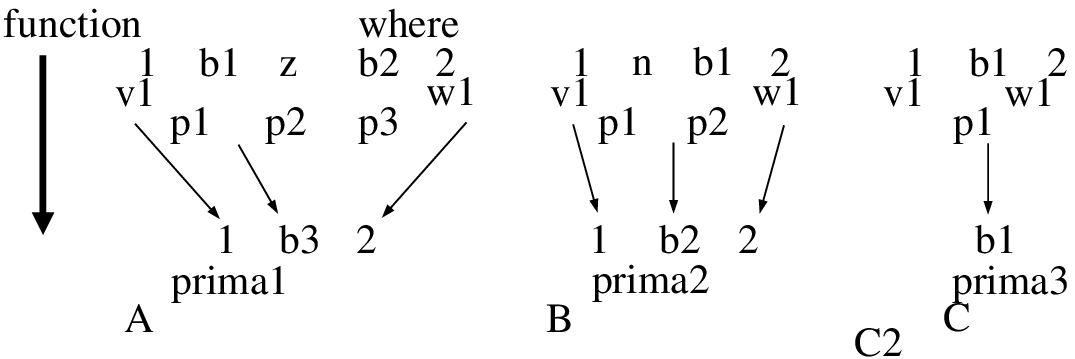,width=\textwidth}
\begin{center}      
 Figure~6.9 
\end{center}
\end{figure}

We describe next how to express A1--A5 above in second-order
logic. See Section~\ref{auxiliary} for the auxiliary formulae used below.\\[0.2cm]
$\mathrm{A1} \equiv \mathrm{LINEAR}(C, E_C)$\\[0.2cm]
$\mathrm{A2} \equiv \mathrm{LINEAR}_2(\mathit{ST}, E_\mathit{ST})$\\[0.2cm]
$\mathrm{A3} \equiv \forall s \, s' \, t_1 \, t_2 \, k_1 \, k_2 \big( \mathrm{A3.1} \wedge  \mathrm{A3.3} \wedge \mathrm{A3.3} \wedge  \mathrm{A3.4}\big)$ where
\begin{itemize}
\item A3.1 expresses ``$M$ is a function,  $M: \mathit{ST} \rightarrow C$''.\\
$\mathrm{A3.1} \equiv ( (M(s, t_1, t_2) \wedge M(s, k_1, k_2)) \rightarrow ( (t_1 = k_1 \wedge t_2 = k_2) \wedge \mathit{ST}(s) \wedge C(t_1, t_2)))$
\item A3.2 expresses ``$M$ is injective''.\\ 
$\mathrm{A3.2} \equiv ( (M(s, k_1, k_2) \wedge M(t_1, k_1, k_2)) \rightarrow s = t_1)$ 
\item A3.3 expresses ``$M$ is total''.\\
$\mathrm{A3.3} \equiv (\mathit{ST}(s) \rightarrow \exists t_1' t_2' (M(s, t_1', t_2')))$
\item A3.4 expresses ``$M$ preserves PATH in $E_\mathit{ST}$ and $E_C$''.\\
$\mathrm{A3.4} \equiv ( (M(s, t_1, t_2) \wedge M(s', k_1, k_2) \wedge \mathrm{PATH}_\mathit{ST}(s, s')) \rightarrow \mathrm{PATH}_{E_C}(t_1, t_2, k_1, k_2))$
\end{itemize}
$\mathrm{A4} \equiv \forall s_1 s_2 \big( (C_\wedge(s_1, s_2) \rightarrow \neg
C_\vee(s_1, s_2)) \wedge (C_\wedge(s_1, s_2) \rightarrow \neg
C_\neg(s_1, s_2)) \wedge$\\
\hspace*{1.8cm}  $(C_\wedge(s_1, s_2) \rightarrow \neg
C_((s_1, s_2)) \wedge (C_\wedge(s_1, s_2) \rightarrow \neg
C_)(s_1, s_2)) \wedge$\\
\hspace*{1.8cm}  $(C_\wedge(s_1, s_2) \rightarrow \neg
C_0(s_1, s_2)) \wedge (C_\wedge(s_1, s_2) \rightarrow \neg
C_1(s_1, s_2)) \wedge \cdots \big) \wedge$\\
\hspace*{0.8cm} $\forall s_1 s_2 \big( C(s_1, s_2) \rightarrow (C_\wedge(s_1, s_2) \vee
C_\vee(s_1, s_2) \vee C_\neg(s_1, s_2) \vee C_((s_1, s_2) \vee C_)(s_1, s_2)
\vee$\\
\hspace*{3.9cm}$C_0(s_1, s_2) \vee C_1(s_1, s_2)) \big) \wedge$\\
\hspace*{0.8cm} $\forall s_1 s_2 \big( (C_\wedge(s_1, s_2) \rightarrow C(s_1, s_2)) \wedge
(C_\vee(s_1, s_2) \rightarrow C(s_1, s_2)) \wedge $\\
\hspace*{1.9cm}$(C_\neg(s_1, s_2)
\rightarrow C(s_1, s_2)) \wedge (C_((s_1, s_2) \rightarrow C(s_1, s_2)) \wedge $ \\
\hspace*{1.9cm}$(C_)(s_1, s_2) \rightarrow C(s_1, s_2)) \wedge (C_0(s_1, s_2) \rightarrow C(s_1, s_2))
\wedge$\\
\hspace*{1.9cm}$(C_1(s_1, s_2) \rightarrow C(s_1, s_2)) \big)$\\[0.3cm]
$\mathrm{A5} \equiv \forall x \big(\mathit{ST}(x) \rightarrow \exists E_v \, f_1 \, f_2 \,
l_1 \, l_2 \, f_1' \, f_2' \, l_1' \, l_2' \big( \alpha_d \vee \alpha_e \vee
(\alpha_0 \wedge (\alpha_a \vee \alpha_b \vee \alpha_c))\big)\big)$ where\\[0.2cm]
The function $E_v$ maps the formula at stage $x$ to the formula at stage
$x+1$. The subformula $\alpha_d$ corresponds to the last transition, i.e., the
transition to the last formula in $(C, E_C)$ (``$0$'' or ``$1$''). The
subformula $\alpha_e$ corresponds to the last formula in $(C, E_C)$. The
subformulae $\alpha_a$, $\alpha_b$ and $\alpha_c$ correspond to the three
possible cases ($a$), ($b$) and ($c$) as in Figure~6.9, according
to which sort of operation is the one involved in the transition from the
formula in stage $x$ to the next formula in $(C, E_C)$. Note that the
transition to the last formula $\alpha_d$ is necessarily an instance of case
($c$) in Figure~6.9. For case ($c$) in Figure~6.9, $E_v$
is not total in its domain, since $(v_1, v_2) \big( ( \big)$ and $(w_1, w_2)
\big( ) \big)$ are not mapped. For the last formula, $E_v$ is not injective,
since $(f_1', f_2') = (l_1', l_2')$ (i.e., $f_1' = l_1'$ and $f_2' =
l_2'$) (see Figure~6.11).\\[0.2cm]
$\alpha_0 \equiv \mathrm{A5.1} \wedge \mathrm{A5.2} \wedge \mathrm{A5.3} \wedge \mathrm{A5.4}$ where 
\begin{itemize}
\item A5.1 expresses ``$x$ is not the leaf in $E_\mathit{ST}$, and it is not the predecessor of the leaf''. \\
$\mathrm{A5.1} \equiv \exists y y_1 (E_\mathit{ST}(x, y) \wedge E_\mathit{ST} (y, y_1) )$
\item A5.2 expresses ``$E_v:C \rightarrow C $ is a partial injection  mapping the formula in $(C,
E_C)$ in stage $x$ to the formula in $(C, E_C)$ in stage $E_\mathit{ST}(x)$''\\
$\mathrm{A5.2} \equiv \forall s_1 s_2 t_1 t_2 k_1 k_2 \big( ( (E_v(s_1, s_2, t_1,
t_2) \wedge E_v(s_1, s_2, k_1, k_2)) \rightarrow$ \\ 
\hspace*{4cm} $( (t_1 = k_1 \wedge t_2 = k_2) \wedge C(s_1, s_2) \wedge C(t_1,
t_2)) ) \wedge$\\
\hspace*{3.3cm} $( (E_v(s_1, s_2, k_1, k_2) \wedge E_v(t_1, t_2, k_1, k_2))
\rightarrow$\\
\hspace*{4cm} $(s_1 = t_1 \wedge s_2 = t_2))\big)$
\item A5.3 expresses ``$( (f_1,f_2),(l_1,l_2))$ and $( (f_1',f_2'),(l_1',l_2'))$ are the \emph{delimiters} of the two formulae as in Figure~6.10''.\\
$\mathrm{A5.3} \equiv M(x, f_1, f_2) \wedge \mathrm{A5.3.1} \wedge \mathrm{A5.3.2} \wedge \mathrm{5.3.3}$ where
\begin{itemize}
\item A5.3.1 expresses ``$M(E_\mathit{ST}(x), E_C(l_1,l_2))$''.
\item A5.3.2 expresses ``$E_C(l_1, l_2) = (f'_1, f'_2)$''.
\item A5.3.3 expresses ``$E^{-1}_C(M(E_\mathit{ST}(E_\mathit{ST}(x))),l_1',l_2')$''.
\end{itemize}
\item A5.4 expresses ``$E_v$ maps nodes from the subgraph induced by $( (f_1, f_2),(l_1,l_2))$ to the subgraph induced by $( (f_1',f_2'),(l_1',l_2'))$''.\\
$\mathrm{A5.4} \equiv \forall y_1 y_2 z_1 z_2 \big( E_v(y_1, y_2, z_1, z_2)
\rightarrow$\\
\hspace*{3.8cm} $\big( \mathrm{PATH}_{E_C}(f_1,f_2,y_1,y_2) \wedge
\mathrm{PATH}_{E_C}(y_1,y_2,l_1,l_2) \wedge$\\ 
\hspace*{4cm} $\mathrm{PATH}_{E_C}(f_1',f_2',z_1,z_2) \wedge
\mathrm{PATH}_{E_C}(z_1,z_2,l_1',l_2') \big) \big) \wedge$\\
\hspace*{1cm} $E_v(f_1, f_2, f_1', f_2') \wedge E_v(l_1, l_2, l_1', l_2')$
\end{itemize} 
\begin{figure}[h!]\label{figure15}
\psfrag{E}{\scriptsize$E_v$}
\psfrag{f1}{\scriptsize$(f_1, f_2)$}
\psfrag{f1b}{\scriptsize$(f_1', f_2')$}
\psfrag{z1}{\scriptsize$(z_{11}, z_{12})$}
\psfrag{z1b}{\scriptsize$(z_{11}', z_{12}')$}
\psfrag{z2}{\scriptsize$(z_{21}, z_{22})$}
\psfrag{z2b}{\scriptsize$(z_{21}', z_{22}')$}
\psfrag{v1}{\scriptsize$(v_1, v_2)$}
\psfrag{v1b}{\scriptsize$(v_1', v_2')$}
\psfrag{l1}{\scriptsize$(l_1, l_2)$}
\psfrag{l1b}{\scriptsize$(l_1', l_2')$}
\psfrag{w1}{\scriptsize$(w_1, w_2)$}
\psfrag{w1b}{\scriptsize$(w_1', w_2')$}
\psfrag{p}{\scriptsize$(p_{11}', p_{12}')$}
\psfrag{text}{\scriptsize Left side of the window}
\psfrag{case1}{\scriptsize Cases (a) and (b)}
\psfrag{case2}{\scriptsize Case (c)}
\psfrag{preserves}{\scriptsize$E_v$ preserves $E_C$}
\centering\epsfig{file=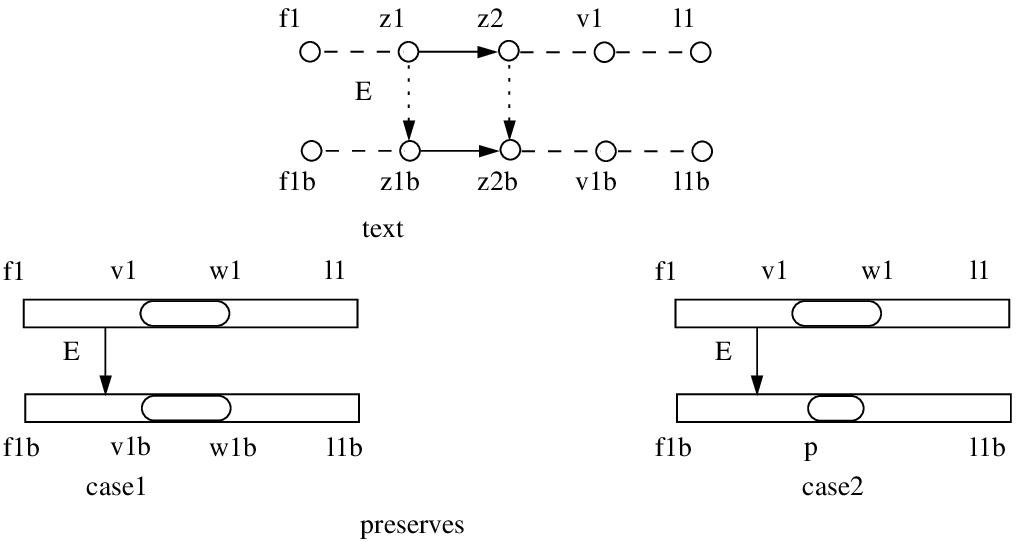,width=\textwidth}
\begin{center}      
 Figure~6.10  
\end{center}
\end{figure}
$\alpha_a \equiv \exists v_1 v_2 w_1 w_2 v_1' v_2' w_1' w_2' p_{11} p_{12}
p_{21} p_{22} p_{31} p_{32} p_{11}' p_{12}' \big($
\begin{flushright}$\mathrm{A5.5} \wedge \mathrm{A5.6} \wedge \mathrm{A5.7} \wedge \mathrm{A5.8} \wedge \mathrm{A5.9} \big)$ where \end{flushright} 
\begin{itemize}
\item A5.5 expresses ``$( (v_1,v_2),(w_1,w_2))$ and $( (v_1',v_2'),(w_1',w_2'))$
define the \emph{window of change}, that is the segment of the formula that is affected (changed) in the transition from stage $x$ to stage $x+1$ of the evaluation (see Cases (a) and (b) in Figure~6.10)''.\\ 
$\mathrm{A5.5} \equiv \mathrm{PATH}_{E_C}(f_1, f_2, v_1, v_2) \wedge 
\mathrm{PATH}_{E_C}(w_1, w_2, l_1, l_2) \wedge E_C(p_{11}, p_{12}, p_{21},
p_{22}) \wedge$\\
\hspace*{1.1cm} $E_C(p_{21}, p_{22}, p_{31}, p_{32}) \wedge E_C(v_1, v_2,
p_{11}, p_{12}) \wedge E_C(p_{31}, p_{32}, w_1, w_2)$\\
\hspace*{1.1cm} $C_((v_1, v_2) \wedge C_)(w_1, w_2) \wedge
\mathrm{PATH}_{E_C}(f_1', f_2', v_1', v_2') \wedge 
\mathrm{PATH}_{E_C}(w_1', w_2', l_1', l_2') \wedge$\\ 
\hspace*{1.1cm} $E_C(v_1',v_2',p_{11}',p_{12}') \wedge
E_C(p_{11}',p_{12}',w_1',w_2') \wedge 
E_v(p_{11},p_{12},p_{11}',p_{12}') \wedge$ \\
\hspace*{1.1cm} $E_v(v_1,v_2,v_1',v_2') \wedge
E_v(w_1, w_2, w_1',w_2') \wedge C_((v_1', v_2') \wedge C_)(w_1', w_2')$
\item A5.6 expresses ``$E_v$ preserves $E_C$ outside of the window of change, and preserves
left and right side of the window of change (see Figure~6.10)''.\\ 
$\mathrm{A5.6} \equiv \forall z_{11} z_{12} z_{21} z_{22} z_{11}' z_{12}' z_{21}'
z_{22}' \big($\\
\hspace*{1.1cm} $\big( (\mathrm{PATH}_{E_C}(f_1,f_2,z_{11},z_{12}) \wedge
\mathrm{PATH}_{E_C}(z_{21}, z_{22}, v_1, v_2) \wedge$\\
\hspace*{1.4cm} $E_C(z_{11}, z_{12}, z_{21}, z_{22}) \wedge E_v(z_{11},
z_{12}, z_{11}', z_{12}') \wedge E_v(z_{21}, z_{22}, z_{21}', z_{22}') )
\rightarrow$\\
\hspace*{1.4cm} $(\mathrm{PATH}_{E_C}(f_1',f_2',z_{11}',z_{12}') \wedge
\mathrm{PATH}_{E_C}(z_{21}', z_{22}', v_1', v_2') \wedge$\\
\hspace*{1.6cm} $E_C(z_{11}',z_{12}',z_{21}',z_{22}')) \big) \wedge$\\
\hspace*{1.1cm} $\big( (\mathrm{PATH}_{E_C}(w_1,w_2,z_{11},z_{12}) \wedge
\mathrm{PATH}_{E_C}(z_{21}, z_{22}, l_1, l_2) \wedge$ \\ 
\hspace*{1.4cm} $E_C(z_{11},z_{12},z_{21},z_{22}) \wedge E_v(z_{11}, z_{12}, z_{11}', z_{12}') \wedge E_v(z_{21},
z_{22}, z_{21}', z_{22}')) \rightarrow$ \\
\hspace*{1.4cm} $(\mathrm{PATH}_{E_C}(w_1',w_2',z_{11}',z_{12}') \wedge
\mathrm{PATH}_{E_C}(z_{21}', z_{22}', l_1', l_2') \wedge$\\
\hspace*{1.6cm} $E_C(z_{11}',z_{12}',z_{21}',z_{22}')) \big) \big)$
\item A5.7 expresses ``$E_v$ preserves symbols in left side of the window of change''.\\
$\mathrm{A5.7} \equiv \forall z_{11} z_{12} z_{11}' z_{12}' \big($\\
\hspace*{1.1cm} $\big(\mathrm{PATH}_{E_C}(f_1,f_2,z_{11},z_{12}) \wedge
\mathrm{PATH}_{E_C}(z_{11}, z_{12}, v_1, v_2) \wedge E_v(z_{11}, z_{12},
z_{11}', z_{12}')\big)$\\
\hspace*{2cm} $\rightarrow \big(\mathrm{PATH}_{E_C}(f_1',f_2',z_{11}',z_{12}') \wedge
\mathrm{PATH}_{E_C}(z_{11}', z_{12}', v_1', v_2') \wedge$\\
\hspace*{2.7cm} $\big( (C_{(}(z_{11},z_{12}) \wedge C_{(}(z_{11}', z_{12}'))
\vee (C_{)}(z_{11},z_{12}) \wedge C_{)}(z_{11}', z_{12}')) \vee$\\
\hspace*{2.9cm} $(C_{\wedge}(z_{11},z_{12}) \wedge C_{\wedge}(z_{11}', z_{12}'))
\vee (C_{\vee}(z_{11},z_{12}) \wedge C_{\vee}(z_{11}', z_{12}')) \vee$\\ 
\hspace*{2.9cm} $(C_{0}(z_{11},z_{12}) \wedge C_{0}(z_{11}', z_{12}'))
\vee (C_{1}(z_{11},z_{12}) \wedge C_{1}(z_{11}', z_{12}')) \vee$\\ 
\hspace*{2.9cm} $(C_{\neg}(z_{11},z_{12}) \wedge C_{\neg}(z_{11}',
z_{12}'))\big)\big)\big)$
\item A5.8 expresses ``$E_v$ preserves symbols in right side of the window of change''.\\
$\mathrm{A5.8} \equiv \forall z_{11} z_{12} z_{11}' z_{12}' \big($\\
\hspace*{1.1cm} $\big(\mathrm{PATH}_{E_C}(w_1,w_2,z_{11},z_{12}) \wedge
\mathrm{PATH}_{E_C}(z_{11}, z_{12}, l_1, l_2) \wedge E_v(z_{11}, z_{12},
z_{11}', z_{12}')\big)$\\
\hspace*{2cm} $\rightarrow \big(\mathrm{PATH}_{E_C}(w_1',w_2',z_{11}',z_{12}') \wedge
\mathrm{PATH}_{E_C}(z_{11}', z_{12}', l_1', l_2') \wedge$\\
\hspace*{2.7cm} $\big( (C_{(}(z_{11},z_{12}) \wedge C_{(}(z_{11}', z_{12}'))
\vee (C_{)}(z_{11},z_{12}) \wedge C_{)}(z_{11}', z_{12}')) \vee$\\
\hspace*{2.9cm} $(C_{\wedge}(z_{11},z_{12}) \wedge C_{\wedge}(z_{11}', z_{12}'))
\vee (C_{\vee}(z_{11},z_{12}) \wedge C_{\vee}(z_{11}', z_{12}')) \vee$\\ 
\hspace*{2.9cm} $(C_{0}(z_{11},z_{12}) \wedge C_{0}(z_{11}', z_{12}'))
\vee (C_{1}(z_{11},z_{12}) \wedge C_{1}(z_{11}', z_{12}')) \vee$\\ 
\hspace*{2.9cm} $(C_{\neg}(z_{11},z_{12}) \wedge C_{\neg}(z_{11}',
z_{12}'))\big)\big)\big)$
\item A5.9 expresses ``In $(p_{11}', p_{12}')$ we get the \emph{result} of applying the operator $\theta$ in $(p_{21}, p_{22})$ to the Boolean values $b_1$, in $(p_{11},p_{12})$, and $b_2$ in $(p_{31}, p_{32})$ (see (a) in Figure~6.9)''.\\
$\mathrm{A5.9} \equiv \big( (C_0(p_{11},p_{12}) \wedge C_0(p_{31}, p_{32}) \wedge
C_\wedge(p_{21},p_{22}) \wedge C_0(p_{11}', p_{12}')) \vee $\\
\hspace*{1.2cm} $(C_0(p_{11},p_{12}) \wedge C_0(p_{31}, p_{32}) \wedge
C_\vee(p_{21},p_{22}) \wedge C_0(p_{11}', p_{12}')) \vee $\\
\hspace*{1.2cm} $(C_0(p_{11},p_{12}) \wedge C_1(p_{31}, p_{32}) \wedge
C_\wedge(p_{21},p_{22}) \wedge C_0(p_{11}', p_{12}')) \vee $\\
\hspace*{1.2cm} $(C_0(p_{11},p_{12}) \wedge C_1(p_{31}, p_{32}) \wedge
C_\vee(p_{21},p_{22}) \wedge C_1(p_{11}', p_{12}')) \vee $\\
\hspace*{1.2cm} $(C_1(p_{11},p_{12}) \wedge C_0(p_{31}, p_{32}) \wedge
C_\wedge(p_{21},p_{22}) \wedge C_0(p_{11}', p_{12}')) \vee $\\
\hspace*{1.2cm} $(C_1(p_{11},p_{12}) \wedge C_0(p_{31}, p_{32}) \wedge
C_\vee(p_{21},p_{22}) \wedge C_1(p_{11}', p_{12}')) \vee $\\
\hspace*{1.2cm} $(C_1(p_{11},p_{12}) \wedge C_1(p_{31}, p_{32}) \wedge
C_\wedge(p_{21},p_{22}) \wedge C_1(p_{11}', p_{12}')) \vee $\\
\hspace*{1.2cm} $(C_1(p_{11},p_{12}) \wedge C_1(p_{31}, p_{32}) \wedge
C_\vee(p_{21},p_{22}) \wedge C_1(p_{11}', p_{12}')) \big)$
\end{itemize}

The subformulae $\alpha_b$ and $\alpha_c$ that correspond to the cases ($b$)
and ($c$) in Figure~6.9, are similar to $\alpha_a$. 
For the clarity of presentation, we omit those formulae. Furthermore, it should be clear how to build
them using $\alpha_a$ as template. Moreover, the complete formulae can be found
in \cite{[Rei11]}. We present next the remaining two subformulae,
namely $\alpha_d$ and $\alpha_e$. \\[0.2cm]
$\alpha_d \equiv \exists y \big( E_\mathit{ST}(x,y) \wedge \neg \exists z
(E_\mathit{ST}(y,z)) \wedge$\\[0.2cm]
\hspace*{0.8cm} $\exists p_{11} p_{12} p_{11}' p_{12}' \big(M(x, f_1, f_2)
\wedge M(y, p_{11}', p_{12}') \wedge$\\[0.2cm]
\hspace*{3.1cm} $E_C(f_1, f_2, p_{11}, p_{12}) \wedge E_C(p_{11}, p_{12}, l_1,
l_2) \wedge E_C(l_1, l_2, p_{11}', p_{12}') \wedge$\\[0.2cm]
\hspace*{3.1cm} $\neg \exists p_{21}' p_{22}'(E_C(p_{11}', p_{12}', p_{21}',
p_{22}')) \wedge$\\[0.2cm]
\hspace*{3.1cm} $C_((f_1,f_2) \wedge C_)(l_1,l_2) \wedge$\\[0.2cm]
\hspace*{3.1cm} $( (C_1(p_{11}, p_{12}) \wedge C_1(p_{11}', p_{12}')) \vee
(C_0(p_{11}, p_{12}) \wedge C_0(p_{11}', p_{12}')) ) \big)\big)$\\[0.2cm]
Note that the first line in $\alpha_d$ expresses ``$x$ is the predecessor of the leaf in $E_\mathit{ST}$'', so that this case corresponds to the last transition (see Figure~6.11). Also note that the last transition is necessarily an instance of case~(c) in Figure~6.9''\\[0.2cm] 
\begin{figure}[h!]\label{figure16}
\psfrag{f}{\scriptsize$(f_1, f_2)$}
\psfrag{p}{\scriptsize$(p_{11}, p_{12})$}
\psfrag{l}{\scriptsize$(l_1, l_2)$}
\psfrag{1}{\scriptsize$($}
\psfrag{b1}{\scriptsize$b_1$}
\psfrag{2}{\scriptsize$)$}
\psfrag{CE}{\scriptsize$(C, E_C)$}
\psfrag{ST}{\scriptsize$(\mathit{ST}, E_{\mathit{ST}})$}
\psfrag{E}{\scriptsize$E_v$}
\psfrag{equation}{\scriptsize$(p_{11}', p_{12}') = (f_1', f_2') = (l_1', l_2')$}
\psfrag{x}{\scriptsize$x$}
\psfrag{y}{\scriptsize$y$}
\psfrag{pb}{\scriptsize$(p_{11}', p_{12}')$}
\psfrag{M}{\scriptsize$M$}
\centering\epsfig{file=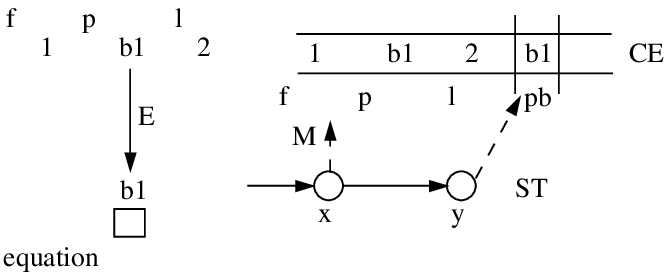,height=4cm}
\begin{center}      
 Figure~6.11  
\end{center}
\end{figure}

\noindent
$\alpha_e \equiv \mathrm{A5.10} \wedge \exists p_1' p_2' \big(M(x, p_1', p_2') \wedge \mathrm{A5.11} \wedge \mathrm{A5.12} \big)$ where
\begin{itemize}
\item A5.10 expresses ``$x$ is the leaf in $E_\mathit{ST}$''.\\
$\mathrm{A5.10} \equiv \neg \exists y (E_\mathit{ST}(x,y))$
\item A5.11 expresses ``$(p_1', p_2')$ is the leaf in $E_C$''.\\
$\mathrm{A5.11} \equiv \neg \exists y_1' y_2' (E_C(p_1',p_2',y_1',y_2'))$
\item A5.12 expresses ``the last formula in $(C, E_C)$ is $1$''.\\ 
$\mathrm{A5.12} \equiv C_1(p_1',p_2')$
\end{itemize}

\subsection{Auxiliary Formulae}\label{auxiliary}

For the sake of completeness, we define next the remaining auxiliary formulae used through the previous
subsections. We assume an edge relation $E$ and a total order~$\leq$.\\[0.2cm]
``$x = 0$'' $\equiv \neg \exists y (y \neq x \wedge y \leq x)$\\[0.2cm]
``$x = 1$'' $\equiv \exists y (y \neq x \wedge y \leq x \wedge \neg \exists z (z \neq x \wedge z \neq y \wedge y \leq z \wedge z \leq
x) \wedge \neg \exists z (z \neq y \wedge z \leq y))$\\[0.2cm]
$\mathrm{SUC}_\leq(x,y) \equiv x \leq y \wedge y \neq x \wedge \neg \exists z (z \neq x \wedge z \neq y \wedge x  \leq z \wedge z \leq
y)$\\[0.2cm]
$\mathrm{PRED}_\leq(y, x) \equiv \mathrm{SUC}_\leq(x, y)$\\[0.2cm] 
$\mathrm{PATH_E}(v,w)$ is used to denote the following formula which is
satisfied by a given graph ${\bf G}$ iff $(v,w)$ is in the transitive closure
of the relation $E^{\bf G}$. \\[0.2cm]
$\mathrm{PATH}_E(v,w) \equiv v = w \, \vee \exists V' E' \big(V'(v) \wedge V'(w) \wedge \mathrm{A1} \wedge \mathrm{A2} \wedge \mathrm{A3} \wedge \mathrm{A4} \wedge \mathrm{A5}\big)$
\begin{itemize}
\item A1 expresses ``$(V', E')$ is a subgraph of $(V, E)$ with no loops''.\\
$\mathrm{A1} \equiv \forall x y ( E'(x, y) \rightarrow (V'(x) \wedge V'(y) \wedge E(x, y))) \wedge \forall x (V'(x) \rightarrow V(x)) \wedge \forall x (\neg E'(x,x))$
\item A2 expresses ``$v$ is the only minimal node''.\\
$\mathrm{A2} \equiv \neg \exists x (E'(x, v)) \wedge \forall y ( (V'(y) \wedge y
\neq v) \rightarrow \exists x (E'(x, y)))$
\item A3 expresses ``$w$ is the only maximal node''.\\
$\mathrm{A3} \equiv \neg \exists x (E'(w, x)) \wedge \forall y ( (V'(y) \wedge y
\neq w) \rightarrow \exists x (E'(y, x)))$
\item A4 expresses ``all nodes except $v$ have input degree $1$''.\\
$\mathrm{A4} \equiv \forall z ( (V'(z) \wedge z \neq v) \rightarrow \exists x (E'(x, z) \wedge \forall y ( (V'(y) \wedge E'(y, z))
\rightarrow y = x)))$
\item A5 expresses ``all nodes except $w$ have output degree $1$''.\\
$\mathrm{A5} \equiv \forall z ( (V'(z) \wedge z \neq w) \rightarrow \exists x (E'(z, x) \wedge \forall y ( (V'(y) \wedge E'(z, y))
\rightarrow y = x)))$
\end{itemize}

That is, $\mathrm{PATH}_E(v,w)$ expresses ``$(V', E')$ is a linear subgraph of $(V, E)$, with minimal node
$v$ and maximal node $w$''. We use a similar 
strategy to define the next auxiliary formula $\mathrm{LINEAR}(V, E)$ which expresses ``$(V, E)$ is a linear graph''.\\[0.2cm]
$\mathrm{LINEAR}(V, E) \equiv \forall x y (\mathrm{PATH}_E(x,y) \vee
\mathrm{PATH}_E(y, x)) \wedge$\\
\hspace*{2.6cm} $(\exists xy (x \neq y) \rightarrow \forall x (\neg E(x,x))) \wedge$\\
\hspace*{2.6cm} $\exists v w \big(V(v) \wedge V(w) \wedge$\\
\hspace*{3.4cm} $\neg \exists x (E(x, v)) \wedge \forall y ( (V(y) \wedge y
\neq v) \rightarrow \exists x (E(x, y))) \wedge$\\
\hspace*{3.4cm} $\neg \exists x (E(w, x)) \wedge \forall y ( (V(y) \wedge y
\neq w) \rightarrow \exists x (E(y, x))) \wedge$\\
\hspace*{3.4cm} $\forall z ( (V(z) \wedge z \neq v) \rightarrow$\\
\hspace*{4cm} $\exists x (E(x, z) \wedge \forall y ( (V(y) \wedge E(y, z))
\rightarrow y = x)))$\\
\hspace*{3.4cm} $\forall z ( (V(z) \wedge z \neq w) \rightarrow$\\
\hspace*{4cm} $\exists x (E(z, x) \wedge \forall y ( (V(y) \wedge E(z, y))
\rightarrow y = x)))\big)$\\[0.2cm]
Note that we only allow loops in a linear graph when it has only one node.

In a similar way we can define the second-order formula $\mathrm{LINEAR}_2(V’,E’)$ where the free second-order variables have arity $2$ and $4$ respectively.

We also use the formula $\mathrm{PATH}_{\mathrm{E_C}}(x_1,x_2,y_1,y_2)$ with free first-order variables $x_1$, $x_2$, $y_1$, $y_2$, where the set of vertices is a binary relation, and the set of edges is a $4$-ary relation (see Figures~6.7 and~6.8).

\section{$\mathrm{SATQBF}$ in Third-Order Logic}

In this section we show how to build a formula in third-order logic that
expresses $\mathrm{SATQBF}$. We omit the tedious details of the subformulae
which can be built following the same patterns than in the detailed exposition of the
second-order formula for $\mathrm{SATQBF}_k$. 

Roughly, we first express the existence of a third-order alternating valuation ${\bf
T}_v$ applicable to a given $\mathrm{QBF}$ formula $\varphi$. Then we proceed
to evaluate the quantifier-free part $\varphi'$ of $\varphi$ on each leaf
valuation ${\bf L}_v$ of ${\bf T}_v$. For this part we use the same
second-order subformulae than for $\mathrm{SATQBF}_k$. That is, from 
$\varphi'$ and ${\bf L}_v$, we build a Boolean sentence $\phi$ on $\{0, 1\}$ by
replacing each occurrence of a Boolean variable $x$ in $\varphi'$ by a constant
$0$ or $1$ according to the Boolean value assigned by ${\bf L}_v$ to $x$, and
then we evaluate $\phi$.  

Unlike the case with $\mathrm{SATQBF}_k$ in which the input formulae all have a same \emph{fixed} number $k$ of alternating blocks of quantifiers, in the case of $\mathrm{SATQBF}$ the number of alternating blocks $k \geq 1$ of quantifiers in the input formulae is \emph{not fixed}. That is, 
we need to take into account that the input formula can have any arbitrary number $k \geq 1$ of alternating blocks of quantifiers. We assume w.l.o.g. that the quantification in the input formula $\varphi$ has the form \\[0.2cm]
$\exists x_{11} \cdots \exists x_{1l_1} \forall x_{21} \cdots \forall
x_{2l_2} \exists x_{31} \cdots \exists x_{3l_3} \cdots Q x_{k1} \cdots Q
x_{kl_k} ($\\[0.2cm]
\hspace*{3.6cm} $\varphi'(x_{11}, \ldots,  x_{1l_1}, x_{21} \ldots, x_{2l_2}, x_{31}, \ldots,
x_{3l_3}, \ldots, x_{k1}, \ldots, x_{kl_k}))$\\[0.2cm]
where $k \geq 1$, the formula
$\varphi'$ is a quantifier-free Boolean formula and $Q$ is $\exists$ if $k$ is
odd, or $\forall$ if $k$ is even. To represent the formulae as relational structures, we use the same encoding based in word models as in Section~\ref{SATQBFk}.

We present a sketch of the third-order formula $\varphi_{\mathrm{SATQBF}}$ that expresses $\mathrm{SATQBF}$. We follow a top-down approach, 
leaving most of the fine details of the formulae in the lowest level of abstraction as an exercise for the reader. 
At the highest level of abstraction, we can think of $\varphi_{\mathrm{SATQBF}}$ as a third-order formula that expresses. 
\[
\text{``There is a third-order alternating valuation ${\bf T}_v$ applicable to
$\varphi$, which satisfies $\varphi$''.} 
\]

At the next level of abstraction we can express $\varphi_{\mathrm{SATQBF}}$ in third-order logic as follows.\\[0.2cm]
$\exists {\cal V}_t \, {\cal E}_t \, {\cal B}_t \, V_t \, E_t \, \Big(\mathrm{A1} \wedge \mathrm{A2} \wedge \mathrm{A3} \wedge \mathrm{A4} \wedge \mathrm{A5}\big)$ where 
\begin{itemize}[leftmargin=*]
\item A1 expresses ``${\cal B}_t: {\cal V}_t \rightarrow \{0, 1\}$''.
\item A2 expresses ``$G_t = (V_t, E_t)$ is a linear graph which represents the sequence of
quantified variables in $\varphi$''.
\item A3 expresses ``$({\cal V}_t, {\cal E}_t)$ is a third-order binary tree with all its leaves at the
same depth, which is in turn equal to the length of $(E_t, V_t)$''.
\item A4 expresses ``$({\cal V}_t, {\cal E}_t, {\cal B}_t)$ is a \emph{third-order alternating
valuation} ${\bf T}_v$ applicable to $\varphi$, i.e., all the nodes in $({\cal
V}_t, {\cal E}_t)$ whose depth correspond to a universally quantified
variable in the prefix of quantifiers of $\varphi$, have exactly one sibling, and its
 value under ${\cal B}_t$ is different than that of the given node, and
all the nodes whose depth correspond to an existentially quantified variable
in the prefix of quantifiers of $\varphi$, are either the root or have no
siblings''.
\item A5 expresses ``Every \emph{leaf valuation} in $({\cal V}_t, {\cal E}_t, {\cal B}_t)$
  satisfies $\varphi'$''.
\end{itemize}
Recall that we use uppercase calligraphic letters for third-order variables and plain uppercase letters for second-order variables. In particular, ${\cal V}_t$, ${\cal E}_t$ and ${\cal B}_t$ are third-order variables while $V_t$ and $E_t$ are second-order variables.

Finally, we describe the strategies to express A2--A5 in third-order logic.

\begin{itemize}
\item[A2.] $\mathrm{LINEAR}(V_t, E_t) \wedge \mathrm{A2.1}$ where
\begin{itemize}
\item A2.1 expresses ``The length of $G_t$ is equal to the number of variables in the prefix of
quantifiers of $\varphi$. That is, there is a relation $V_p$ which is a partial bijection from the quantifier prefix of $\varphi$
(restricted to the $X$'s in the quantifier prefix) to $V_t$, which maps every
$X$ in the quantifier prefix to its corresponding node in $G_t$, and which preserves $E_t$
and $\leq^{ {\bf G}_\varphi}$ in $G_t$ and $\varphi$ (restricted to the $X$'s in the
quantifier prefix), respectively''.\\
See (A) in Subsection~\ref{exparti} for more details.
\end{itemize}

\item[A3.] Let ${\cal E}_t\restriction_{{\cal S}_d}$ denote the restriction of the third-order relation ${\cal
E}_t$ to the nodes in the third-order set ${\cal S}_d$. We can express A3 as follows:\\
$\mathrm{A3.1} \wedge \mathrm{A3.2} \wedge \mathrm{A3.3} \wedge \mathrm{A3.4}$ where
\begin{itemize}
\item A3.1 expresses ``$({\cal V}_t, {\cal E}_t)$ is a third-order connected graph that has one root and one or
more leaves''.
\item A3.2 expresses ``Except for the root node, all nodes in $({\cal V}_t, {\cal E}_t)$ have input
degree~$1$''. 
\item A3.3 expresses ``Except for the leaf nodes, all nodes in $({\cal V}_t, {\cal E}_t)$ have
output degree $1$ or $2$''.
\item A3.4 expresses ``All leaf nodes in $({\cal V}_t, {\cal E}_t)$ have the same
depth, which is in turn equal to the length of $(V_t, E_t)$''
\end{itemize}
A3.1, A3.2 and A3.3 can be expressed in third-order logic as follows:\\
$\exists R \big( \forall Z ( {\cal V}_t(Z) \rightarrow \mathrm{PATH}_{{\cal
E}_t}(R, Z)) \wedge$\\
\hspace*{0.5cm} $\neg \exists S_1 ({\cal E}_t(S_1, R)) \wedge$\\
\hspace*{0.5cm} $\exists S_1 (\neg \exists S_2 ( {\cal E}_t(S_1, S_2)))
\wedge$\\
\hspace*{0.5cm} $\forall Z ( ({\cal V}_t(Z) \wedge Z \neq R) \rightarrow
\exists S_1 ({\cal E}_t(S_1, Z) \wedge \forall S_2 ({\cal E}_t(S_2, Z) \rightarrow
S_1 = S_2)))\big) \wedge$\\
$\forall Z \big( {\cal V}_t(Z) \rightarrow \neg \exists S_1 S_2 S_3 (S_1 \neq S_2
\wedge S_2 \neq S_3 \wedge S_1 \neq S_3 \wedge {\cal E}_t(Z, S_1) \wedge {\cal
E}_t(Z, S_2) \wedge$\\
\hspace*{3.6cm} ${\cal E}_t(Z, S_3))\big)$\\[0.1cm]
Regarding A3.4, we can express it as follows:\\
$\forall X \big( \mathrm{A3.4.1} \rightarrow \big(\exists {\cal S}_d \, {\cal D} (\mathrm{A3.4.2} \wedge \mathrm{A3.4.3} \wedge {\cal S}_d(X) \wedge \mathrm{A3.4.4}\wedge \mathrm{A3.4.5} ) \big) \big)$ where
\begin{itemize} 
\item A3.4.1 expresses ``$X$ is a leaf node in $({\cal V}_t, {\cal E}_t)$''.
\item A3.4.2 expresses ``${\cal S}_d \subseteq {\cal V}_t$''.
\item A3.4.3 expresses ``${\cal D} : V_t \rightarrow {\cal S}_d$ is a bijection that
preserves $E_t$ and ${\cal E}_t\restriction_{{\cal S}_d}$''.
\item A3.4.4 expresses ``${\cal D}^{-1}(X)$ is the leaf node in $G_t = (V_t, E_t)$''.
\item A3.4.5 expresses ``${\cal S}_d$ includes the root of $({\cal V}_t, {\cal E}_t)$''.
\end{itemize}
\item[A4.] We can express A4 as follows (refer to Figures~6.2 and~6.5):\\
$\mathrm{A4.1} \wedge \forall x \forall {\cal S}_d \big( (V_t(x) \wedge \mathrm{A4.2} \wedge \mathrm{A4.3} ) \rightarrow$\\
\hspace*{2.2cm} $\forall {\cal D} ( (\mathrm{A4.4} \wedge \mathrm{A4.5} ) \rightarrow ( ( \mathrm{A4.6} \rightarrow \mathrm{A4.7}) \wedge (\mathrm{A4.8} \rightarrow \mathrm{A4.9} )))\big)$ where 
\begin{itemize}
\item A4.1 expresses ``${\cal B}_t$ is a total function from ${\cal V}_t$ to $\{0, 1\}$''.
\item A4.2 expresses ``${\cal S}_d \subseteq {\cal V}_t$''.
\item A4.3 expresses ``$({\cal S}_d, {\cal E}_t\restriction_{ {\cal S}_d})$ is a linear graph which includes the root of $({\cal V}_t, {\cal E}_t)$''.
\item A4.4 expresses ``${\cal D}$ is a bijection from the initial subgraph of $G_t$ up to $x$, to ${\cal S}_d$''.
\item A4.5 expresses ``${\cal D}$ preserves $E_t$ and ${\cal E}_t\restriction_{ {\cal S}_d}$''.
\item A4.6 expresses ``the predecessor of $V^{-1}_p(x)$ in $\leq^{ {\bf G}_\varphi}$ is $\forall$''.
\item A4.7 expresses ``${\cal D}(x)$ has exactly one sibling in $({\cal V}_t, {\cal E}_t)$ and ${\cal B}_t$ of that sibling is not equal to ${\cal B}_t({\cal D}(x))$''.
\item A4.8 expresses ``the predecessor of  $V^{-1}_p(x)$ in $\leq^{ {\bf G}_\varphi}$ is $\exists$''.
\item A4.9 expresses ``${\cal D}(x)$ has no siblings in $({\cal V}_t, {\cal E}_t)$, or ${\cal D}(x)$ is the root in $({\cal V}_t, {\cal E}_t)$''.
\end{itemize}
\item[A5.] $\forall {\cal S}_v \big( ( \mathrm{A5.1} \wedge \mathrm{A5.2} )\rightarrow \exists {\cal D} \, B_t \big( \mathrm{A5.3} \wedge \mathrm{A5.4} \wedge \mathrm{A5.5} \big)\big)$ where 
\begin{itemize}
\item A5.1 expresses ``${\cal S}_v \subseteq {\cal V}_t$''. 
\item A5.2 expresses ``$({\cal S}_v, {\cal E}_t\restriction_{ {\cal S}_v})$ is a linear graph which includes the root and a leaf of $({\cal V}_t, {\cal E}_t)$''.
\item A5.3 expresses ``${\cal D}$ is a bijection from $V_t$ to ${\cal S}_v$ which preserves $E_t$ and ${\cal E}_t\restriction_{ {\cal S}_v}$''.
\item A5.4 expresses ``$B_t$ is a total function from $V_t$ to $\{0, 1\}$ which coincides with ${\cal B}_t({\cal S}_v)$ w.r.t. ${\cal D}$''.
\item A5.5 expresses  ``the leaf valuation represented by $(V_t, E_t, B_t)$ satisfies the quantifier-free subformula $\varphi'$ of $\varphi$''.
\end{itemize}
Note that, A5.5 can be expressed as in Subsection~\ref{part_ii}.
\end{itemize}

\begin{remark}
Note that while in the third-order formulae in~A4 and~A5 we have used universal third-order quantification (for ${\cal S}_d$ and ${\cal D}$ in~A4, and for ${\cal S}_v$  in~A5), it is not actually needed, and existential third-order quantification is enough. These are the only sub-formulae where we have used universal third-order quantification. Hence, we strongly believe that our third-order formula can be translated in a rather technical way into an existential third-order formula.

Let us consider the sketch for an existential third-order formula equivalent to the formula in~A4 (the existential formula for~A5 is easier). We can say that for every node $x$ in the graph $(V_t,E_t)$, and for every set $Z$ that is a node in the third-order graph $({\cal V}_t, {\cal E}_t)$, and such that there is a third-order set ${\cal S}_d$ of nodes in the third-order graph $({\cal V}_t, {\cal E}_t)$, such that the restriction of the edge relation ${\cal E}_t$ to the third-order set ${\cal S}_d$, together with ${\cal S}_d$, form a (third-order) subgraph that is a linear graph whose root is the root of the third-order graph $({\cal V}_t, {\cal E}_t)$, and whose leaf is the set $Z$, and such that its length is the length of the initial subgraph of the graph $(V_t,E_t)$, up to the node $x$, if the variable represented by $x$ in the input formula $\varphi$ is universally quantified, then the node $Z$ in the third-order graph $({\cal V}_t, {\cal E}_t)$ has exactly one sibling in that graph, and that sibling has a different value assigned by ${\cal B}_t$ than the value assigned by ${\cal B}_t$ to $Z$. On the other hand, if the variable represented by $x$ in the input formula $\varphi$ is existentially quantified, then the node $Z$ in the third-order graph $({\cal V}_t, {\cal E}_t)$ has no sibling in that graph.
To say that ``the third order graph induced by the set ${\cal S}_d$ in the graph $({\cal V}_t, {\cal E}_t)$, whose leaf is the set $Z$, has the same length as the initial subgraph of the graph $(V_t,E_t)$, up to the node $x$'', we say that there is a binary third-order relation $\cal D$ which is a bijection between the set of nodes in the initial subgraph of the graph $(V_t,E_t)$, up to the node $x$, and the third-order set ${\cal S}_d$, and which preserves $E_t$ and the restriction of the edge relation ${\cal E}_t$ to the third-order set ${\cal S}_d$.
\end{remark}

\section{Final Considerations}

Let $\exists\mathrm{SO}^{\leq 2}$ denote the restriction of $\exists$SO to formulae with second-order
variables of arity $\leq 2$. 
As pointed out in \cite{[DLS98]}, it is open whether on graphs full $\exists$SO is strictly more expressive than $\exists\mathrm{SO}^{\leq 2}$. Also as pointed out in \cite{[DLS98]}, no concrete example of a graph property in PSPACE that is not in binary NP has been found yet, even though it is known that such properties exist. 
Hence, it would be worthwhile to find an example of a
PSPACE query on graphs that cannot be expressed in
$\exists\mathrm{SO}^{\leq 2}$. The gained experience
on writing non-trivial queries in second-order logic, can prove to be 
a valuable platform to make progress on these kind of open problems. In 
particular, we used a second-order variable of arity 4 in Section 6.
We used it to represent (together with other variables) a linear digraph which, for each of the
leaf valuations, encodes a
sequence of word models corresponding to the different stages of 
evaluation of the quantifier free part
of the input QBF$_k$ formula. Since
the size of the Boolean formula in each stage is linear in the size of the input
 QBF$_k$ formula, and the number of connectives in the formula is also
linear, the length of the complete sequence of Boolean formulae is
quadratic. Therefore, we conjecture that arity $4$ is actually a lower bound,   
though we have not attempted to prove it yet. 
In general, the exploration of properties which force us to work 
with intermediate structures of size greater than linear w.r.t. the input, 
seems a reasonable way of approaching these kind of open problems.

As noted earlier, there are second-order queries that are difficult to express in the language of second-order logic, 
but which have an elegant and simple characterization in third-order logic. Therefore it would be interesting to explore possible characterizations of fragments of third-order logic that admit translations of their formulae to equivalent formulae in second-order logic. 
This way, those fragments of third-order logic could be assimilated to high-level programming languages, while second-order logic would be the corresponding low-level programming language. In turn, this would allow us to express complex second-order queries with greater abstraction of the low-level details, thus minimizing the probability of error.

\bibliographystyle{plain}
\bibliography{biblio.bib}

\end{paper}
\end{document}